\documentclass[iop]{emulateapj}
\usepackage{amsmath}
\usepackage{graphicx}
\usepackage{epstopdf}
\usepackage{hyperref}
\usepackage[usenames,dvipsnames]{color}  %SdM: allows fancy color names
\usepackage{natbib}
\newcommand{\kms}{\ensuremath{\,\rm{km}\,\rm{s}^{-1}}}
\newcommand{\Msun}{\ensuremath{\,M_\odot}}
\newcommand{\Rsun}{\ensuremath{\,R_\odot}}
\newcommand{\Lsun}{\ensuremath{\,L_\odot}}
\newcommand{\changes}[1]{{ {#1}}}

\newcommand{\vrot}{\ensuremath{\,v_{\rm rot}}}
\newcommand{\vsini}{\ensuremath{\,v\,{\sin i}}}
\newcommand{\fbin}{\ensuremath{\,f_{\rm bin}}}

\newcommand{\dd}{\mathrm{d}}

\shorttitle{The rotation rates of massive stars and the role of binary interaction}
\shortauthors{De Mink et al.}

\begin{document}
\title{The rotation rates of massive stars: \\ the role of binary interaction through tides, mass transfer and mergers}

\author{S.E.~de Mink\altaffilmark{1,2*}  N.~Langer\altaffilmark{3}, R.G.~Izzard\altaffilmark{3}, H. Sana\altaffilmark{4} \& A. de Koter\altaffilmark{4,5,6} } 

\altaffiltext{1}{Space Telescope Science Institute, Baltimore, Md, USA}
\altaffiltext{2}{Johns Hopkins University, Baltimore, Md, USA}
\altaffiltext{3}{Argelander-Institut f\"ur Astronomie der Universit\"at Bonn, Germany}
\altaffiltext{4}{Astronomical Institute Anton Pannekoek, University of Amsterdam, The Netherlands}
\altaffiltext{5}{Utrecht University, The Netherlands}
\altaffiltext{6}{Institute of Astronomy, KU Leuven, Belgium}
\altaffiltext{*}{Hubble Fellow.}

\begin{abstract}

Rotation is thought to be a major factor in the evolution of massive stars -- especially at low metallicity -- with consequences for their chemical yields, ionizing flux and final fate.  Deriving the birth spin distribution is of high priority given its importance as a constraint on theories of massive star formation and as input for models of stellar populations in the local Universe and at high redshift. Recently, it has become clear that the majority of massive stars interact with a binary companion before they die. We investigate how this affects the distribution of rotation rates, through stellar winds, expansion, tides, mass transfer and mergers.

For this purpose, we simulate a massive binary-star population typical for our Galaxy assuming continuous star formation. We find that, because of binary interaction, $20_{-10}^{+5}$\% of all massive main-sequence stars have projected rotational velocities in excess of $200\kms$. We evaluate the effect of uncertain input distributions and physical processes and conclude that the main uncertainties are the mass transfer efficiency and the possible effect of magnetic braking, especially if magnetic fields are generated or amplified during mass accretion and stellar mergers.

The fraction of rapid rotators we derive is similar to that observed. If indeed mass transfer and mergers are the main cause for rapid rotation in massive stars, little room remains for rapidly rotating stars that are born single. This implies that spin down during star formation is even more efficient than previously thought. In addition, this raises questions about the interpretation of the surface abundances of rapidly rotating stars as evidence for rotational mixing. Furthermore, our results allow for the possibility that all early-type Be stars result from binary interactions and suggest that evidence for rotation in explosions, such as long gamma-ray bursts, points to a binary origin.

\end{abstract}


%%%%%%%%%%%
%%%%%%%%%%%  1. Introduction  
%%%%%%%%%%%

\section{Introduction}

The origins of the distribution of initial stellar masses and initial stellar rotation rates is not yet
fully understood \citep[e.g.][]{Krumholz2011,Rosen+2012}. For the latter it is a key issue that 
stars inherit spin from their parental molecular clouds, whose specific angular momentum is 
orders of magnitude larger than what can be placed in a rotating star \citep[e.g.][]{Bodenheimer1995}. 
Based on the observations of low-mass pre-main sequence and massive main-sequence stars,
it appears that most stars reach the zero-age main sequence rotating significantly 
below their maximum possible rotation speed \citep[e.g.][]{Hartmann+1989, Huang+2010}. 
How angular momentum is expelled during the star formation process constitutes
a classic problem \citep{Mestel1965, McKee+2007,Krumholz+2009, 
Larson2010}. Observed rotation rates of stars on the zero-age main sequence provide 
important constraints for the theory of star formation. Unfortunately, massive stars become 
visible only once core hydrogen burning is already well underway
\citep{Yorke1986} such that their initial rotational velocity distribution is not observationally accessible.

The embedding of massive stars in the early phases of their lives is indeed unfortunate as the
initial rotation rate is thought to be a fundamental parameter determining their evolutionary fate, comparable to their initial mass and metallicity. Evolutionary models predict that rotation has major consequences for the 
core hydrogen-burning phase \citep{Maeder+2000a, Heger+2000a, 
Hirschi+2004, Yoon+2005, Brott+2011a, Potter+2012a, Ekstrom+2012}. Furthermore, there is growing evidence that
in a fraction of massive stars, the final collapse and explosion is governed by
rapid rotation, giving rise to hyper-energetic supernovae and long-duration gamma-ray bursts 
\citep{Yoon+2006, Woosley+2006, Georgy+2009}.

Several processes may affect the stellar rotation rates of massive main-sequence stars after they
are born, causing deviations from the initial values. 
First, magnetic braking may cause them to spin down. In low-mass main-sequence stars
magnetic breaking is ubiquitous, due to a dynamo process operating in their convective envelopes
\citep{Skumanich1972, Soderblom+1993}. The fraction of more massive stars that
show evidence for a magnetic field appears, however, to be smaller than about 15\% \citep{Donati+2009, Wade+2012b}

Second, during core hydrogen burning massive stars expand by about a factor
of three. It appears, however, that a corresponding spin down of the surface layers is
prevented by an analogous contraction of the stellar core and by efficient transport of angular
momentum from the core to the envelope \citep{Ekstrom+2008, Brott+2011},
independent of the detailed treatment of the internal angular momentum transport processes.
The effects of main-sequence expansion on the surface rotation rates thus seem modest as well.

Third, even non-magnetic massive stars lose angular momentum due to their winds 
 \citep{Langer1998}. 
The winds of massive main-sequence stars,
which are fairly well understood \citep{Kudritzki+2000, Mokiem+2007a}, are stronger at higher 
mass and metallicity \citep{Vink+2001} and for more rapid rotation \citep{Friend+1986}.
While for the majority of massive stars this wind induced angular momentum loss can be
neglected --- for example, stars of LMC composition below 14\,M$_{\odot}$ lose less than 10\% of
their angular momentum this way \citep{Langer2012} --- it can be important for
O-stars and rapid rotators.

Finally,  ---likely most important for the population of massive stars as a whole--- are changes in the stellar spin
due to close binary interaction. In recent decades, it has become evident that the majority of massive stars are 
formed in close binary systems \citep[][]{Mason+1998, Mason+2009, Sana+2011, Sana+2012a}. 
In these systems, the spin rate of both components can be drastically affected, by tides 
\citep[e.g.][]{Zahn1975, Hut1981, de-Mink+2009a}, mass transfer \citep[e.g.]{Packet1981, Pols+1991, 
Petrovic+2005, Dervisoglu+2010}, or when the two stars merge \citep[e.g.][]{Podsiadlowski+1992,Tylenda+2011}. 
\citet[]{Sana+2012} showed that about two-thirds of all massive main-sequence O-stars
are expected to strongly interact with a companion. In the majority of
cases, after such interaction, either the binary is not expected to be recognized as such due to a large luminosity ratio and a large period, 
or there is no binary any more due to merging or the break-up of the binary 
as consequence of the supernova explosion of one star \citep{de-Mink+2011,de-Mink+2012-letter}. 
Consequently, any massive star population that is not extremely 
young contains a considerable number of apparently single stars whose spins are strongly altered 
by previous binary evolution.

In this paper we focus on the effects of binarity on the evolution of the rotation rates of a population
of massive main-sequence stars. We do so by constructing binary population synthesis models that
account for the latter three effects: the changing moment of inertia, wind losses, and --- most
importantly --- binary interaction. It is our aim, after presenting our method and assessing the uncertainties, 
to compare our results with observed distributions of rotational velocities of massive
main-sequence stars \citep[e.g.][P.~L. Dufton et al. 2012, submitted, O. Ram\'irez-Agudelo et al. 2012, in 
preparation]{Mokiem+2006, Daflon+2007, Wolff+2008,Hunter+2008, Penny+2009, Huang+2010}. This will provide the 
first quantitative assessment of the difference between the initial rotational velocity distribution
and the observed, present-day distributions.

%%%%%%%%%%%
%%%%%%%%%%%  2. Code  
%%%%%%%%%%%

\section{Code} 
\label{code}

We employ a rapid binary evolutionary code \emph{binary\_c}
\citep{Izzard+2004,Izzard+2006,Izzard+2009} that makes use of fitting formulae
\citep{Hurley+2000} to stellar models \citep{Pols+1998} to describe
the structure of single stars as they evolve as a function of their mass,
age and metallicity.  To study the effects of binary interaction by
tides and winds and through mass transfer and mergers  we use the evolutionary algorithms originally developed by \citet{Tout+1997} and \citet{Hurley+2002}. These
algorithms approximate the properties and evolution of a star after
mass loss or mass accretion by switching to the prescriptions of a star with the appropriate mass, core mass and relative age. These approximations enable us to follow the evolution of a
particular binary system up to the remnant stage in less than a CPU
second, which is needed to compute extensive grids of models that span
the multi-dimensional parameter space characteristic for binary
systems. In addition it allows us to explore the effects of uncertain parameters
\citep[cf. ][]{Izzard+2009}.  For the purpose of this study we updated and extended various aspects of this code, which are described in the Appendix.  A brief overview is given in the remainder of this section. 

Throughout this paper, we present results in terms of the \emph{`rotational
velocity'} with which we refer to the tangential velocity due to
rotation at the equator of the star, $v_{\rm rot}$.  
We account for the flattening of a rotating star assuming Roche geometry.   \changes{ In this approximation, the ratio of the radius at the equator and the polar radius becomes as large as  $R_{\rm eq} / R_{\rm p} = 3/2$ when the star approaches} the Keplerian limit, which is given by  
 \begin{equation}
\Omega_{\rm K} = \sqrt{ G M / R_{\rm eq, K}^3},\label{omegaK}
\end{equation}
where $G$ refers to the gravitational constant, $M$ is the stellar mass and $R_{\rm eq, K}$ is the equatorial radius of the star that rotates at the Keplerian limit.   Equivalently, the  Keplerian rotational velocity is given by 
\begin{equation}
v_{\rm K} = \Omega_{\rm K}  R_{\rm eq, K}
\end{equation}
 
\changes {To model deformation by rotation we use the fact that the polar radius is hardly affected by rotation.  Even when the star rotates at the Keplerian limit the polar radius deviates by less than
2\% from its non-rotating value for stars with masses between
3-20\Msun~and it deviates by less than 5\% for stars with masses
between 1-60\Msun~\citep{Ekstrom+2008}.
This property allows us to use the stellar radii $R_*$ in our code, which are
based on non-rotating stellar models, as a good approximation for the
polar radii of rotating stars, $R_{\rm p} (\omega) \approx R_{\rm p}
(0) = R_*$, where we use the shorthand notation $\omega \equiv \Omega / \Omega_{\rm K}$. The equatorial radii can then be obtained by computing the shape of the Rochelobe as a function of the rotation rate, as is further explained in the appendix. }
 To account for gravity darkening in stars rotating near their Keplerian limit, we assume that the maximum rotational velocity that can be observed is limited to a fraction $f_{\rm dark}$ of the Keplerian rotation rate. Based on \citet{Townsend+2004} we adopt $f_{\rm dark} = 0.7$. Although this treatment is very simple, it reproduces the detailed
simulations of  \citet{Townsend+2004} accurately enough for our purpose.  Varying this parameter affects the extend of the high velocity tail in our distribution of rotation rates, but it does not have a significant on our main predictions.

\paragraph {Processes affecting all stars}
We account for mass loss via stellar winds using prescriptions by  \citet{Nieuwenhuijzen+1990} and \citet{Vink+2001} as described in \citet{Brott+2011}.  We treat the enhancement of stellar winds due to rotation as in \citet{Maeder+2000}. As justified by \citet{Georgy+2011}, we assume mass loss through  stellar winds to be spherical when computing the associated angular momentum loss. Details can be found in the appendix. \changes{Our models do not account for effects of magnetic fields on angular momentum loss by stellar winds \citep[e.g.][]{Ud-Doula+2009}. We discuss the possible effects  in Section~\ref{sec:assumptions}. } 

As the star evolves along the main sequence, the outer layers expand while the core contracts. We account for changes in the moment of inertia using fitting formulae based on models by \citet{Pols+1998}, see appendix.  The internal rotational profile is approximated assuming rigid rotation, which is a reasonable approximation for main-sequence stars \citep[e.g.][]{Brott+2011}.

\paragraph {Processes affecting stars in close binaries}
We include the effect of tides on the stellar spins and the stellar orbit \citep{Zahn1977, Hurley+2002} and the transfer of angular momentum during mass transfer via an accretion disk or direct impact of the accretion stream onto the surface \citep{Ulrich+1976, Packet1981} as detailed in the Appendix.    We assume that the initial stellar spins are aligned with the orbit.  
One of the main uncertainties in the treatment of binary interaction is the efficiency of mass transfer \citep[e.g]{de-Mink+2007}. In particular, it is unclear how much stars can accrete after reaching the Keplerian limit.  In our standard simulations we follow \citet{Paczynski1991}  and \citet{Popham+1991}  who argue that the accretion disk regulates the mass and angular momentum flux through viscous coupling allowing the star to continue to accrete. We do however limit the accretion rate by the thermal rate of the accreting star as described in the Appendix and in \citet{Hurley+2002}.  In section~\ref{beta} we discuss the effect of this assumption. 
 To account for systems that come in contact during the mass transfer phase we consider a critical mass ratio, $q_{\rm crit}$, as motivated in the Appendix. We assume that the stars come in contact when  $M_{\rm acc}/M_{\rm don} < q_{\rm crit}$, where $M_{\rm don}$ is the mass of the Roche lobe filling star  and $M_{\rm acc}$ is the mass of the companion.  In our standard simulations we adopt $q_{\rm crit, MS} = 0.65$ if the donor star is a main-sequence star, $q_{\rm crit, HG} = 0.4$ when the donor fills it Roche lobe when crossing the Hertz sprung gap.  For more evolved donor we follow \citet{Hurley+2002}.  The effects of these assumptions are discussed in Sect.~\ref{contact}. 
We assume that a merger product efficiently loses the excess angular momentum, such that it rotates at near Keplerian rotation when it has settled to its thermal equilibrium structure \citep[][]{Sills+2005, Glebbeek+2009}.  To account for mixing of fresh hydrogen into the central regions for accreting stars and mergers we follow \citet{Hurley+2002}.  We assume that the convective core size adapts to the new mass and we describe this process using improved prescriptions for the effective mass of the convective core \citep{Glebbeek+2008a}, see Appendix.  To be able to investigate the effect of the uncertainties related stellar mergers we implement a  parametrized prescription that allows us to explore extreme assumptions.   In our standard simulations we assume that a fraction  $\mu_{\rm loss} = 0.1$ of the total system mass is lost during the merger and that  a small fraction $\mu_{\rm mix} = 0.1$ of the envelope is mixed into the convective core, see Appendix.  The effect of these assumptions is discussed in Section~\ref{sec:assumptions}.

\subsection {Simulating a stellar population}
\label{pop}
To investigate the effect of binary interaction on the distribution of
rotation rates of early-type stars, we
simulate a stellar population including binaries under the assumption
of continuous star formation.
We approximate the distribution of the initial masses of single stars and
the primary stars of binary systems by 
\begin{equation}
f_{M_1}(M_1) \dd M_1 \propto M_1^{-\alpha} \dd M_1,
\end{equation}
where $M_1$ denotes the mass of the primary star and $\alpha=2.35\pm0.7$ \citep{Salpeter1955, Kroupa2001}.  Even though one can argue whether an
initial mass function that was derived for single stars can be applied
to binary systems we note that the effect of this assumption on our results is small, see Section~\ref{sec:assumptions}.
 
For the initial distribution of mass ratios, orbital periods and the
corresponding binary fraction we use the distributions by \citet{Sana+2012} based on
extensive monitoring campaigns of O stars in nearby young clusters 
\citep{De-Becker+2006,Hillwig+2006,Sana+2008,Sana+2009,Rauw+2009,Sana+2011a}.
After correcting for biases, the authors find a binary fraction of {$69\pm9$\%} for systems with mass ratios between 0.1
and 1 and orbital periods between $10^{0.15}$ and $10^{3.5}$ days.  The distribution of orbital periods is described by
\begin{equation}
  f_p (P)\, \dd \! \log P   \propto \left( \log P \right)^{\pi}\, \dd \!\log P,    
\end{equation}
where $\pi = -0.55\pm +0.2$. 
The distribution of mass ratios is described by
\begin{equation}
f_q (q)\,\dd q \propto q^{\kappa} \,\dd q, 
\end{equation}
where $q = M_2/M_1$, $M_2$ denotes the mass of the secondary star and $\kappa =
-0.1\pm 0.6$. 
Because these distributions are derived for a very young population, we assume that they well approximate the
initial distribution functions of binary parameters. 
\changes{In our standard simulations we adopt $\alpha=2.35$, $\pi=-0.5$ and
$\kappa=0$. The effects of varying these parameters is discussed in Section~\ref{sec:assumptions}, see also Table~\ref{tab1}.}

We do not include the effect of binaries with companions outside the range of
orbital periods and mass ratios covered by \citet{Sana+2012} because the fraction of such systems is poorly
constrained.  As a result we may underestimate the effect of binaries.  
Systems with mass ratios more extreme than 0.1 may produce rapid rotators
of early type if they interact during the main sequence of the
primary and merge.   

To reduce the dimensions of the parameter space we assume circular orbits.  \citet{Hurley+2002} show that tides circularize the binary orbit just before the onset of mass transfer, implying that the effect of the
initial eccentricity on the further evolution is small.  By not considering the
effect of very wide systems with high eccentricities we may 
underestimate the fraction of stars that interact. 

The initial rotation rate of stars in a binary system does significantly affect the evolution of the binary system, except in extreme and rare cases \citep[e.g.][]{de-Mink+2009a}. The reason is that the stellar spins at birth typically contribute at most a few percent to the total angular momentum budget of the system. 

Since the rotational velocity distribution of massive stars at the zero-age main sequence is poorly constrained
(cf. Sect.~1), we choose a very simple one that allows us to clearly demonstrate the effects of binary interaction.  
We adopt a uniform distribution \changes{for $v_{\rm rot}$ in the range $0-200\kms$, effectively assuming that stars are born with low to moderate rotation rates.}  While the upper limit of $200\kms$ is not physically motivated, it will allow us to unambiguously identify the binary contribution to the formation of rapid rotators.   
Since the initial stellar spins are negligible compared to the amount of angular momentum exchanged during mass transfer
or merger, they do not affect the rotation rates of stars after these interactions. 
Whereas in future work, it may be considered to adapt the initial spin distribution to reproduce the properties 
of suitable observed samples, this is beyond our present scope.   
To compute the distribution of projected rotation rates $v \sin i$, where $i$ is the inclination angle of the binary system we assume that the orientation of the binary orbits are random in space 

Unless an observational campaign is designed to detect binaries, many companions stars will remain undetected.  
We assume that only the rotation rate of the brightest star is measured in this case. Therefore, when constructing
the simulated distribution of rotation rates we only include the most luminous main-sequence star of each binary system.

For our standard simulation we adopt a metallicity of $Z = 0.008$, appropriate for the Large Magellanic Cloud. This metallicity is  also considered representative for star forming regions at a redshift 1--2, at the peak of star formation in the Universe.   The effect of metallicity is discussed in Sect.~\ref{metallicity}.

We derive the distribution of rotation rates for systems that are brighter than $10^4$ and $10^5\Lsun$, respectively.  To put this in perspective,  in our models $10^4$\Lsun~corresponds to the luminosity of a 8.5--12\Msun~main-sequence star, depending on whether we take the model at zero-age or at the end of the main sequence.  Similarly, a luminosity cut-off of $10^5\Lsun$ corresponds to stars with masses in excess of 20--28\Msun.   \changes {Effectively, the first group is dominated by early B-type stars and the second group by O-type stars}   We refrain at this stage from applying other cut-offs such as criteria based on temperature, color or spectral type, since our predictions for the temperatures are less reliable than those for the luminosities.

%%%%%%%%%%%
%%%%%%%%%%%  3. Results 1 Explanation  
%%%%%%%%%%%

\section{The evolution of the rotational velocity for individual systems}

\begin{figure}[t]\center
  \includegraphics[width=0.5\textwidth]{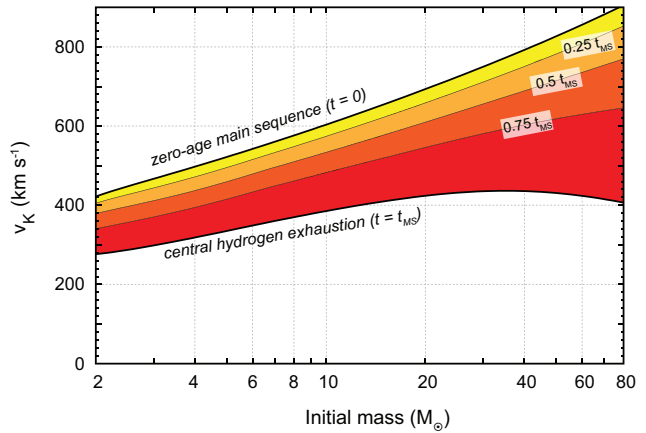}
  \caption{ The evolution of the Keplerian rotational velocity for stars of different initial masses during their main-sequence evolution. Labels show the relative age as a fraction of the main-sequence lifetime $t_{\rm MS}$. This diagram is constructed using stellar radii by \citet{Pols+1998} for a metallicity of $Z=0.008$, using the prescription by \citet{Hurley+2000}. We note that the time-dependent radii for stars more massive than about 40\Msun~ are very uncertain. In the most massive stars the decrease of the Keplerian velocity with time may well be more severe than shown here.  See Section~\ref{remarks} for a discussion. \label{vcrit_vs_mass}}
\end{figure}

As a star evolves, its rotational velocity is affected by various
processes, for example as a result of angular momentum loss through stellar winds. 
While a single star can only lose angular momentum as a result of mass loss, a star in a binary system may either lose or gain angular momentum as it interacts with its companion.  Even when angular momentum is conserved, the rotational velocity of a star
can alter as a result of changes in the stellar
interior.  These effects are discussed in
Section~\ref{remarks}. The effect of binary interaction is discussed in Section~\ref{examples} and~\ref{logp}.

\begin{figure*}[t]\center
  \includegraphics[angle=0, width=\textwidth]{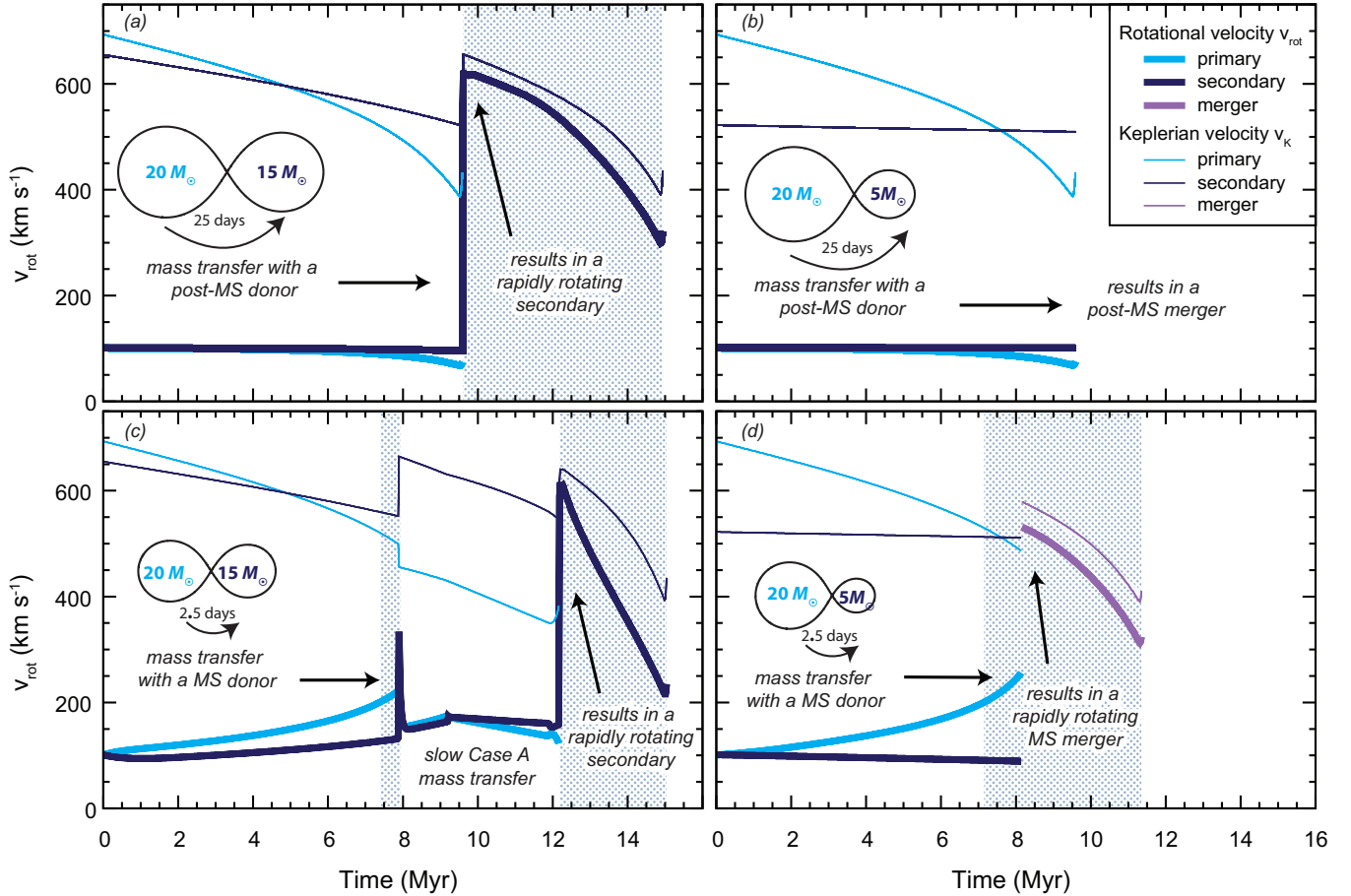}
  \caption{ Four annotated examples of the effect of binary interaction on the stellar rotation rates of main-sequence stars for systems with different initial mass ratios (left versus right panels) and different initial orbital periods (upper versus lower panels) for $Z=0.008$.   Each panel shows the evolution of the equatorial rotational
    velocity $v_{\rm rot}$ of the primary and secondary star (thick lines) as well as the Keplerian rotational velocity $v_{\rm K}$ (thin lines) as long as the stars are on the main sequence.
 Shading highlights the phases during which one
    of the stars is rotating more rapidly than 200\kms.  See
    Sect~\ref{examples} for more information.  \label{vrot_examples}}
\end{figure*}

\subsection {Effect of changes in the stellar structure}
\label {remarks}
Stars expand over the course of their main-sequence evolution. Although one might expect intuitively that the rotational velocity decreases as the star expands, in practice this is not the case. During the main-sequence, contraction and consequent spin-up of the core counteract the effect of the moderate expansion of the envelope. 
In this context it is instructive to investigate how the moment of inertia, $I \sim k R^2$, changes as the star expands.  Here, $k$ denotes the square of the effective gyration radius, which depends on the internal density profile.  If we approximate $k \sim R^{-\xi}$, or equivalently  
\begin {equation}
\xi \equiv - \frac{d \ln k}{d \ln R},
\end{equation}
we find that the exponent $\xi$ varies only slightly during the main sequence with typical values of 1.5--2.  In the case of rigid rotation, we can now
express how the rotation rate $\Omega$, the rotational velocity \vrot~and the ratio of the 
rotation rate and the Keplerian rate $\omega/\omega_{\rm K}$
change as the star expands,
\begin{eqnarray}
  \frac{d \ln \Omega}{d \ln R}  &=& \xi -2 \label{omega}, \\
  \frac{d \ln  \vrot}{d \ln R}  &=& \xi -1\label{vrot},\\
  \frac{d \ln    \Omega/\Omega_{\rm K}  }{d \ln R}   &=&\xi- \frac{1}{2}.\label{omcrit}
\end{eqnarray} 
In other words, given typical values of $\xi$, we find that the decreasing gyration radius compensates for the expansion of the star such that the rotation rate $\Omega$
decreases only slightly as the star evolves (Eq.~\ref{omega}), since $d \ln \Omega / {d \ln R} \approx -0.5\text{--}0$.
The rotational velocity at the equator, \vrot,  increases slightly (Eq.~\ref{vrot}).

Most interestingly, the last expression shows that stars naturally evolve towards the Keplerian limit
(Eq.~\ref{omcrit}), if the amount of angular momentum loss is small and internal angular momentum transport between the core and envelope is efficient.   An important implication of this is that when a star reaches the Keplerian rotation rate, it remains rotating near the Keplerian limit in the absence of an efficient angular momentum loss mechanism. We refer to the excellent discussion by \citet{Ekstrom+2008}, which describes this effect in detailed models of single stars that allow for differential rotation.

Figure~\ref{vcrit_vs_mass} shows the Keplerian rotational velocity $v_{\rm K}$ as a function of the initial mass of a star at different stages during the main sequence.  In zero-age main-sequence stars the Keplerian velocity increases monotonically with initial stellar mass.  The Keplerian velocity drops as stars evolve and expand. The
largest change occurs during the final stages of main-sequence
evolution.  The change in radius of more massive stars during the main
sequence is larger, resulting in a more significant drop in the
Keplerian velocity.  As a result the Keplerian rotational velocity at the end of the main sequence is around 400\kms~with only a weak dependence on the stellar mass.  \changes{Note that the projected rotational velocity for such a star accounting for gravity darkening is considerably smaller, by a factor of  $f_{\rm dark} \langle \sin i \rangle \approx 0.55 $.}

We note that the amount by which the massive stars expand over the main
sequence is not well constrained. The models on which this diagram is
based \citep{Pols+1998} have been calibrated to the radii of eclipsing binaries of
intermediate mass \citep{Pols+1997,Schroder+1997}.  In these models the expansion of massive stars over the course of the main sequence is limited to a
factor of 2--3.  In contrast, the models by \citet{Brott+2011}, \changes{which adopt a larger value for the overshooting parameter,}  predict expansion by a
factor 3--5 for stars in the mass range 5--30\Msun.   While the predictions based on models with both codes show excellent agreement at zero-age, the Keplerian velocities at the end of the main sequence are smaller by about a factor of two 
in the \citet{Brott+2011} models.

Furthermore we note that in the most luminous stars the effects of radiation pressure can not be ignored and the Keplerian limit should be considered as an upper limit to the physical maximum.

\subsection {Examples of the spin evolution of binary
  systems \label{examples}}
In Fig.~\ref{vrot_examples} we depict four examples of the evolution
of the rotational velocity for
main-sequence stars in a binary system.  In each example we assumed an initial mass for the primary star of 20\Msun, a metallicity of $Z=0.008$ and initial rotational velocities of 100\kms~for both stars.  
In the top row we show systems with initial orbital
periods of $P=25$~days.  These systems are so wide that the primary
star fills its Roche lobe only after it leaves the main sequence
as it expands during its hydrogen shell burning phase.  In the bottom
row we assumed an initial orbital period of $P=2.5$~days. In these
tight systems the primary star fills its Roche lobe as a result of 
expansion on the main sequence.  

The panels on the left show examples
in which the initial mass of the secondary is comparable to that of
the primary, $M_2/M_1 = 0.75$.  In these examples one or more phases
of mass transfer eventually lead to spin-up of the companion star.  In the
panels on the right we adopted a more extreme initial mass ratio,
$M_2/M_1 = 0.25$.  In these systems the onset of mass transfer brings
the stars into contact and they merge.  Only in the short period case are the two stars that merge
both still main-sequence stars.  After the merged star regains its
thermal equilibrium it is expected to continue to burn 
hydrogen in the center.

The phases during which one of the stars is rotating more rapidly than
200\kms~are highlighted in Figure~\ref{vrot_examples} with grey
shading.  Notice that during the major part of this phase the
rapidly rotating star is single or appears to be single. The clearest example is shown in panel~(d) where the rapidly rotating star is the product of a 
merger between the two stars.  In panels~(a) and (b) the
rapidly rotating star is the spun up secondary.  The primary star has
lost its hydrogen envelope and is hard to detect, as a
result of its reduced mass, its low luminosity and the wide orbit.
When the primary star finishes its nuclear burning and
explodes it is likely to disrupt the system, leaving the rapidly rotating
secondary behind as a single star.

Besides the drastic effects of mass transfer and coalescence,
Fig.~\ref{vrot_examples} illustrates the effects of other processes
that have a more subtle effect of the stellar rotation rates, which we describe below. 

\paragraph {Changes in the stellar structure} 
The effect of changes on the stellar structure as the star evolves can be observed,
for example, in panel~(a) of Figure~\ref{vrot_examples}. During the first 9.5 Myr
of the evolution of this system both stars reside well within
their Roche lobe and their evolution is similar to that of single stars.
The rotational velocity of both stars remains roughly constant during this phase (cf. Sect.~\ref{remarks}).  When the primary star approaches the end of its main-sequence life time, its expansion accelerates. This leads to the decrease of the rotational velocity that is visible in panel~(a) at an age of 8-9,5~Myr. The expansion is also responsible for the decrease of the Keplerian rotational velocity with time. 

\paragraph {Stellar wind}
The effect of angular momentum loss by stellar winds is small during the major part of the main sequence in these examples. The winds become stronger toward the end of the main sequence, which  together with the expansion, contributes to the decrease in the rotational velocity of the primary discussed above.   However, spin down by winds does play a significant role for the massive, rapidly rotating stars that can be produced as a result of mass transfer.  This effect can be seen most clearly in panel~(c) for ages of 12-15~Myr.  The secondary quickly spins down reducing its rotational velocity by about a factor of three.

\paragraph {Tides}
Tidal interaction tends to synchronize the rotation of the stars with the orbit in systems where the separation between the stars is comparable to the stellar radii.  The systems depicted in the upper and lower panels have initial separation in the order of 
120\Rsun~and 40\Rsun~respectively, slightly less for the systems on the right due to the smaller mass ratios. Tides do not play a significant role during the main sequence of the stars in the upper panels, but they are responsible for the gradual increase in rotational velocity that can be seen in the lower panels for the primary star during the first 8 Myr.  The orbital period remains nearly constant during this phase. The primary star is kept in corotation with the orbit as it expands, which implies that the rotational velocity gradually increases.   In this example, the primary expands by just over a factor of two before it fills its Roche lobe, resulting in a rotational velocity in excess of 200\kms. 
 
In panel~(c) of Figure~\ref{vrot_examples} tides are also important for the
secondary star, during the first mass transfer phase, which starts at
about 8 Myr.  The secondary star spins up as it accretes mass and
angular momentum. However, the tides quickly spin the star down to
synchronous rotation.  This system experiences a mass transfer
phase, which last almost 4~Myr, during which both stars remain in synchronous rotation while the orbit gradually widens.  Around 12~Myr a second rapid phase of mass transfer sets in, i.e. case AB,  as the primary star leaves the main
sequence and expands during hydrogen shell burning.  As a result of the high mass transfer rate and the fact that the orbit widens, tides are no longer effective in preventing the accreting star from spinning up.

\begin{figure*}[t]\center
  \includegraphics[ width=\textwidth]{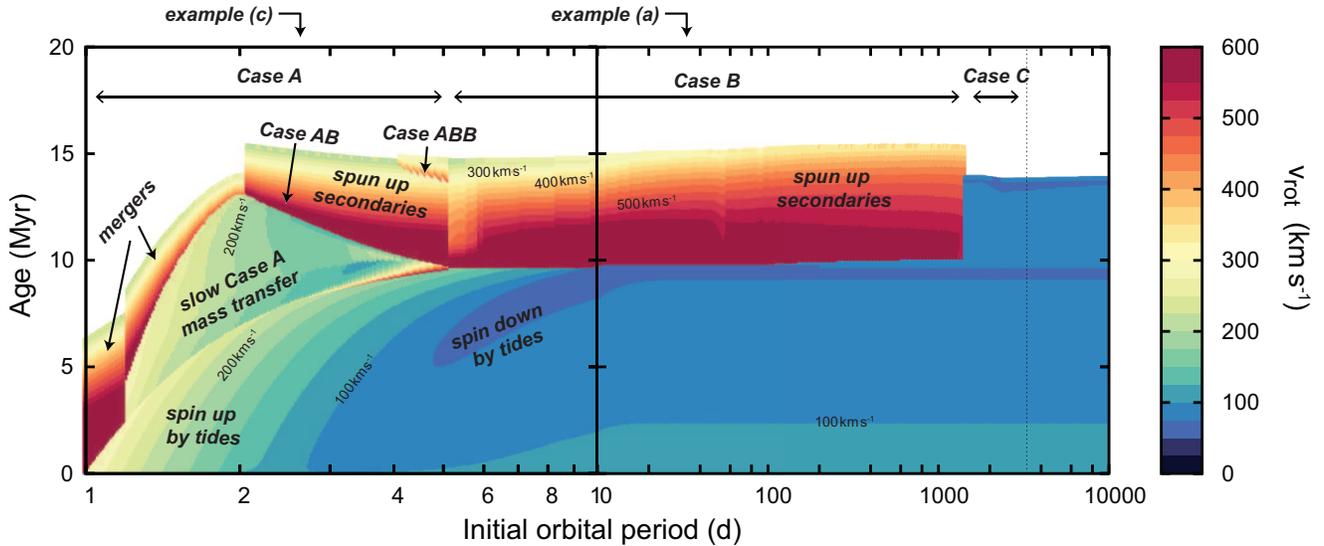}
  \caption{ Rotational velocity (color shading) as a function of
    initial orbital period and time for the brightest main-sequence
    star in a binary system.  We adopt initial masses of 20 and 15\Msun, an
    initial rotational velocities of 100\kms~and a metallicity of
    $Z=0.008$.  As the stars evolve along the main
    sequence their rotational velocity is altered by stellar winds,
    internal evolution, tides and most notably mass accretion.
    The vertical dotted line indicates the maximum separation for which
    this system interacts by mass transfer.   The examples shown in panel (a) and (c) of Fig.~\ref{vrot_examples} part of this simulation. \label{vrot_logP}}
\end{figure*}

\subsection {Effect of the initial separation and mass ratio} \label
{logp}

To further illustrate the effect of the initial binary parameters, we
depict in Figure~\ref{vrot_logP} the evolution of the rotational
velocity of the brightest main-sequence star in a binary system as a
function of the initial orbital period.  For this example we adopted
an initial rotation rate of 100\kms, a metallicity of $Z=0.008$ and
initial masses of 20 and 15\Msun.  The color shading indicates the
equatorial rotational velocity of the brightest main-sequence star in
each system.  Initially this is the primary star, but after mass
transfer the secondary becomes the brightest.

\begin{figure*}[t]\center
  \includegraphics[width=\textwidth]{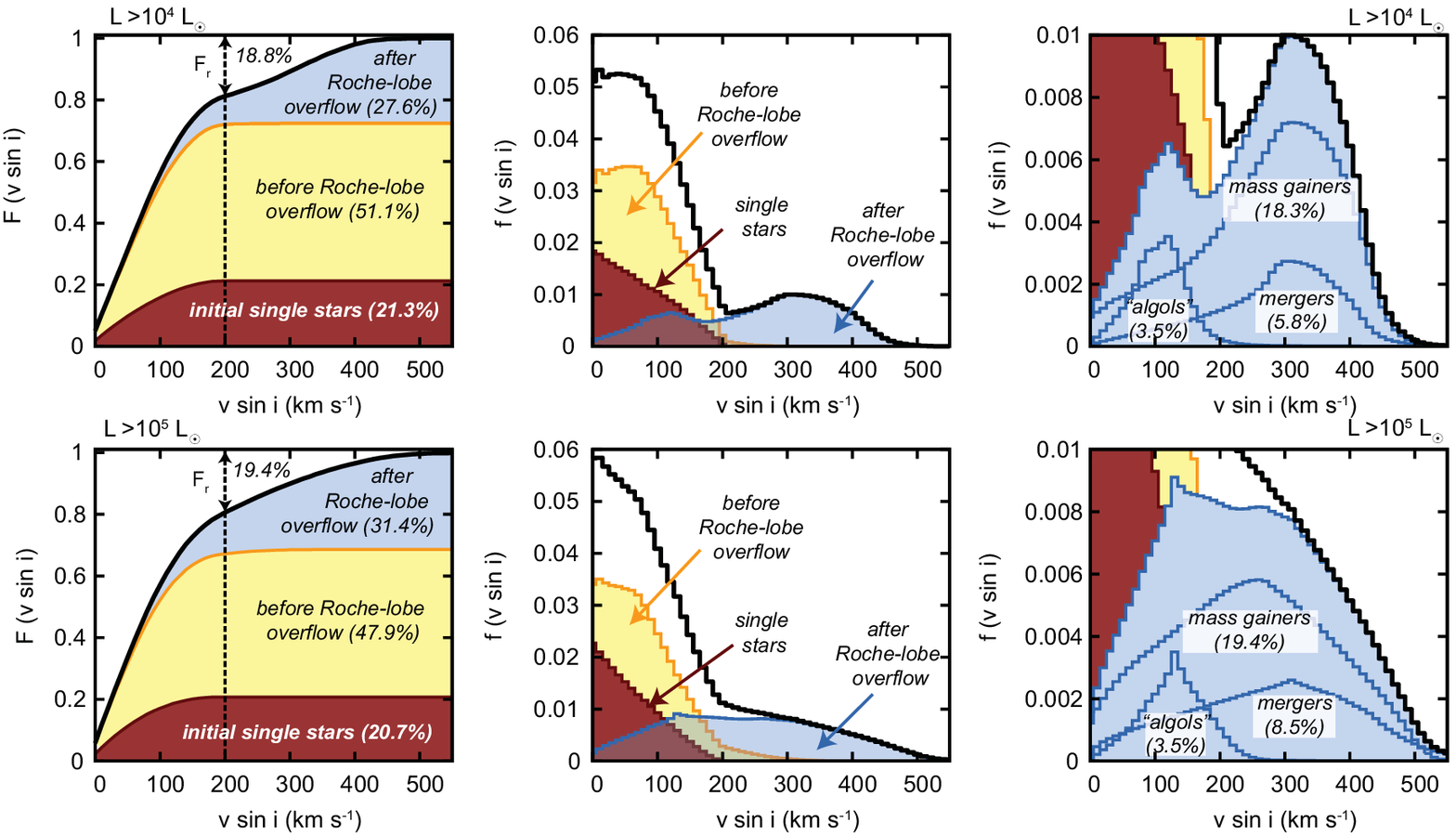}
  \caption{Projected rotation rate distribution for a population of main-sequence stars assuming continuous star formation.   The top and bottom row show results including only stars or systems brighter than  $10^4$ and $10^5\Lsun$.   The left panels show the cumulative distribution function indicating the fraction of stars with projected rotational velocities larger than 200\kms, $F_r$. The center and right panels show the full distribution function and a zoom-in highlighting the contribution of the various binary products. See~Sect.\ref{sec:distribution} for further explanation. \label{vrot_dist}}
\end{figure*}

\paragraph {Mass transfer with a main-sequence donor (Case A)}
In short period systems the expansion of the primary star during its
main-sequence evolution is sufficient to make it fill its Roche
lobe. These systems experience a phase of slow mass transfer that can last for several Myr.  Mass transfer typically occurs via direct impact onto the surface of the secondary.   Tides keep both stars synchronized, which prevents the secondary from reaching very high rotation rates.  This phase ends when the two stars come into
contact and merge (for periods $P \lesssim$~2 days) or when the primary star
leaves the main sequence (for periods P $\approx$~2--5 days) triggering a new mass transfer phase (Case AB). Both cases are expected to lead to the formation of a massive rapidly rotating star.
In certain cases the rapidly rotating star experiences an additional
spin up phase, when the primary fills its Roche lobe again as it expands during He shell burning (Case ABB).  This effect is visible in Figure~\ref{vrot_logP} around 14
Myr for systems with orbital periods of 4--5 days.

The rapidly rotating stars resulting from such close binaries are
typically more massive than the original primary.  As a result of their high mass and large luminosity, their remaining life time is short compared to the rapid rotators
resulting from wider systems and they experience more efficient spin
down by their stellar winds, as can be seen in Figure~\ref{vrot_logP}.

The type of evolution described in this section is representative
for systems with comparable initial masses for the primary and
secondary star.  In systems with more extreme mass ratios, the
secondary star is predicted to swell up and fill its Roche lobe as
well directly after the onset of mass transfer. If the two stars
evolve into a contact situation it is expected that mass is lost
via the outer Lagrangian point, effectively draining angular momentum
from the system. This drives the stars into deeper contact.  We assume
that such systems will merge to form a rapidly rotating single star.
 
\paragraph{Mass transfer with a post-main sequence donor (Case B)}
Systems wider than about 5 days interact when the primary star expands after core hydrogen exhaustion as it crosses the Hertzsprung-Russell diagram on its way to become a red super giant.  This happens during hydrogen-shell burning (for periods P $\approx 5-100$ days) or after the ignition of helium (for periods P $\approx
100-1000$ days)\footnote{Note that part of this group is indicated as Case C in the classical classification scheme \citep{Kippenhahn+1967, Lauterborn1970}, which is based on models for lower mass stars at solar metallicity that reach  giant dimensions before helium ignition.  In our example helium is ignited at smaller radii. Since this group behaves similar to Case B systems we discuss them in this section.}.  The period ranges quoted here are only rough indications, they depend on the mass of the donor star and on metallicity.

Mass accretion in these systems occurs typically via an accretion disk. The companion is quickly spun up approaching the Keplerian velocity after accreting only a few percent of its own mass.  According to  \citet{Popham+1991} mass accretion continues after the star has reached Keplerian rotation as viscous processes transport excess angular momentum outward. The outer edge of the disk is truncated by tides and feeds the excess angular momentum back to the orbit.
 
The mass transfer rate is higher for the wider systems, in which the
donor is more evolved.  In our standard simulations, the mass accretion rate is
limited by the thermal rate, $M_* / \tau_{\rm KH}$, of the accreting star, where $\tau_{\rm KH}$ is the Kelvin Helmholtz timescale.  This is independent of the separation.  As a result we find the general trend that tighter systems evolve more conservatively resulting in more massive secondaries in comparison with wider systems.  The
more massive secondaries have stronger stellar winds which cause them
to spin down more quickly in comparison to less massive secondaries.  This
effect can be seen in the steepness of the color gradient in Figure~\ref{vrot_logP}. Compare for example the change of rotation rate with time for a rapid rotator resulting from a binary with an initial orbital period of about 5.5 days with the rapid rotators resulting from wider systems.  

Surprisingly, the remaining main-sequence lifetime of the secondary is
roughly independent of the amount of accreted mass or orbital
period. This is a consequence of two counteracting effects. The
higher stellar mass implies shorter evolutionary time scales, i.e a
reduction of the remaining life time. At the same time mass accretion
results in mixing of fresh hydrogen into the central regions,
extending the remaining life time.

Systems with extreme mass ratios are expected to enter a phase of
common envelope evolution soon after the onset of mass transfer.
The tighter systems result in a merger, but the product is a
post main-sequence star, either a blue or red supergiant. Although such mergers are very
 interesting, in this paper we focus on main-sequence stars, so
we ignore these in the further discussion. In wider systems where the binding energy of the envelope of the primary is smaller and the momentum and energy in the orbit are
larger, the envelope can be ejected. This process is so fast that it
is not expected to significantly affect the secondary star.  So, we assume
its rotation rate is unaffected.  The only exceptions are
systems that are just wide enough to avoid a merger, but in which the
ejection of the envelope results in significant shrinking of
the orbit, such that tides lock the orbit of the secondary to
the orbit of the naked core of the primary star.   Such systems experience a second phase of mass transfer when the primary expands during helium shell burning.  However, these systems are rare. They do not significantly contribute to the population of rapidly rotating stars resulting from binary evolution.

\paragraph{Mass transfer with supergiant donor (Case C)}
In the widest binaries, $P \approx$~1500--3000 days in our simulations, the donor star develops a convective envelope before it fills its Roche lobe.  When it loses mass we assume it reacts
by expanding, resulting in very high mass transfer rates \citep[see however][]{Woods+2011}.  As the Roche lobe typically shrinks, this leads to a run-away situation
resulting in a common envelope phase.  In our simulations these systems eject the envelope, leaving the secondary star and its spin relatively unaffected.

A common envelope phase is expected for a very wide range of mass
ratios.  Only systems with initial mass ratios close to one can avoid this
phase, see for example \citet{Claeys+2011}.

%%%%%%%%%%%
%%%%%%%%%%%  4. Results: DIstribution  
%%%%%%%%%%%

\section {The distribution of projected rotational velocities}
\label{sec:distribution}

We simulate the distribution of projected rotational velocities for a population of main-sequence stars consisting of single stars and stars in binary systems by assuming continuous star formation and a uniform distribution of low to intermediate rotational velocities at birth $\vrot < 200$\kms, as described in Sect.~\ref{code}. The central panels of Figure~\ref{vrot_dist} show the resulting distributions, $f (\vsini)$, using a bin size $\Delta \vsini = 10\kms$. The upper and lower panels show our results for brightness cut-offs of $10^4$ and  $10^5\Lsun$, respectively (Sect.~\ref{code}). The corresponding cumulative distributions,  
\begin{equation}
F (\vsini) = \sum_0^{\vsini} \!f (\vsini)\,\Delta \vsini
\end{equation}  
are shown in the panels on the left.  

For a luminosity limit of $10^4\Lsun$, the resulting $\vsini$-distribution is bimodal. The majority of the stars have low to intermediate rotational velocities, whereas nearly one fifth has rotational velocities in excess of $200\kms$, i.e. larger than the maximum of the assumed distribution at birth.    The first group consists predominantly of single stars and members of binary systems that have not interacted through Roche-lobe overflow.   The rapid rotators consist almost exclusively of stars that are the product of Roche-lobe overflow.   The top right panel of Figure~\ref{vrot_dist} shows that this group consists mainly of stars that gained mass and angular momentum by mass transfer.  A smaller number are stars merger products, i.e stars result from the coalescence of two main-sequence stars in a contact binary.  

For stars above $10^5\Lsun$, the shape of the distribution changes as can been seen in the two central panels.  The distribution shown in the upper panel shows two separated components.  Instead, the rapid rotators shown in the lower panel form a tail or plateau that merges with the main component at lower \vsini.  This difference is the result of angular momentum loss through stellar winds, which is stronger for the brighter stars.  Even though the brightest stars are included in both distributions, the distribution in the upper panel is dominated by the less bright stars, since the initial mass function favors lower mass stars. 

A further difference between the upper and lower plot in the central panel of Fig.~4 is that the bright star distribution extends to higher rotational velocities.  The reason is that many binary products rotate near their Keplerian limit, which is larger for the brighter, more massive stars (cf., Fig.~1). 

Not all binary products have large projected rotational velocities.  Apart from rapid rotators observed at small inclinations, i.e. near pole-on,  there is a small contribution from systems that are undergoing mass transfer, indicated with the shorthand ``Algols'' in Figure~\ref{vrot_dist}.  This group consists almost entirely of short-period systems that are undergoing a slow, i.e. nuclear timescale, Case~A mass transfer (cf. Sect.~3). The rotational velocity reflects the rotation rate of the accreting star, which has become the brightest star in the system.  Even though this star is accreting mass and angular momentum, tides are efficient enough,  during this slow phase of mass transfer, to prevent it from spinning rapidly.  These systems are easily detected as binaries.  As the donor star fills its Roche lobe these systems are likely to show eclipses or at least ellipsoidal variations.  Examples of such systems are the 30 semi-detached systems in the sample of double-lined eclipsing binaries by \citet{Hilditch+2005}.  

The distribution of the rotational velocities of single stars closely resembles the adopted uniform birth distribution.   The apparent bias towards low rotational velocities is only partially due to spin down by stellar winds and changes in the stellar structure.  The main reason is the projection effect caused by the random distribution of spin axis of the stars.

The rotational velocities of stars in binary systems that have not yet interacted by Roche-lobe overflow can be affected by tides.  The effect of spin up by tides can be seen in the yellow shaded area as a tail of stars rotating between 200 and about 300\kms, which is most pronounced in the lower central panel.   Tides are also responsible 
for the flattening of the yellow curve for pre Roche-lobe overflow systems below 100\kms, as
they counteract the spin down towards the end of the main sequence which is imposed by stellar winds and stellar expansion.    Even though tides may also lead to spin down, they do not produce very slow rotators, since slow synchronized rotation implies wide orbits and tides are no longer effective in wide binaries. 

  We have included mergers separately in Figure~\ref{vrot_dist}, to show how the distribution may change if mergers behave differently than we have assumed in our simulations (see Sect.~\ref{sec:assumptions-physics} for further discussion).

\subsection {Metallicity dependence} \label{metallicity}
\begin{figure}[t]\center
  \includegraphics[width=0.5\textwidth]{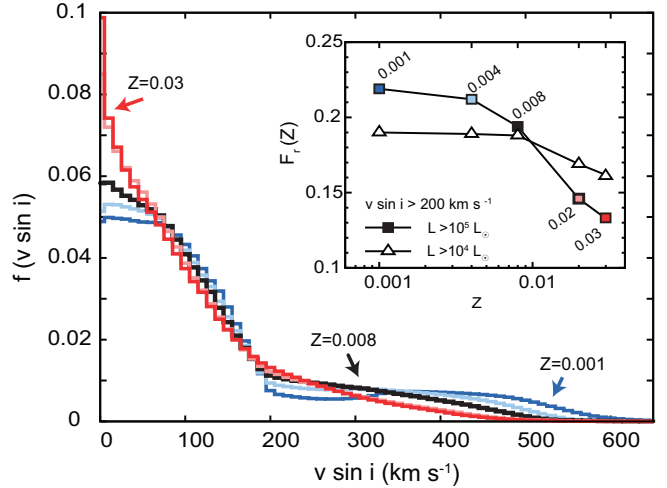}
  \caption{ The distribution of projected rotational velocities $f(\vsini)$ for different metallicities ranging from $Z=0.001$ (blue line) to $0.03$ (red line) for systems brighter than $L > 10^5\Lsun$ containing at least one main-sequence star.  The inset gives the color code and summarizes the effect of metallicity by showing the fraction of stars with projected rotational velocities larger than 200\kms, $F_r$, i.e.. See Sect.~\ref{metallicity} for further information.  \label{Zdep}}
\end{figure}

The metallicity affects the distribution of rotational velocities
via its effect on the stellar wind and on the stellar structure.  In Figure~\ref{Zdep} we show the normalized distribution of rotational velocities for main-sequence stars brighter than $10^5$\Lsun~for a range of metallicities. We assume here that the initial mass function, the binary fraction and the binary distribution functions are not dependent on metallicity. 

At the lowest metallicities shown, the distribution consists of two distinct components with broad peaks at
$\vsini \sim 100$ and $400\kms$. It resembles the distribution shown in the top panels of Figure~\ref{vrot_dist}  for $10^4$\Lsun, although the low metallicity distribution extends to velocities as high as 600\kms. At the highest metallicity the two components merge to one smooth distribution with a strong peak at $\vsini\simeq 0\kms$.  

The inset in  Figure~\ref{Zdep} shows the fraction of stars with projected rotational velocities in excess of 200\kms, $F_r$, as a function of metallicity, Z. 
This fraction is in the range 13--22\%, with larger values for smaller metallicities, and a stronger metallicity
dependence for more massive stars.  
These trends are mainly the result of the dependence of the stellar winds on metallicity and brightness,
and the corresponding wind-induced angular momentum loss.  

Additionally, main-sequence stars tend to be more compact at lower metallicity, where hydrogen burning via the CNO cycle operates less efficiently.  More compact stars have higher Keplerian rotational velocities for the same mass.  This is reflected in the extent of the high velocity tail in Figure~\ref{Zdep} which increases with decreasing metallicity. 
Furthermore, the smaller radii imply somewhat more compact binary systems, which raises the importance
of tidal interactions at low metallicity. 
  
%%%%%%%%%%%
%%%%%%%%%%%  5. Assumptions 
%%%%%%%%%%%

\section {Effects of uncertainties in the adopted assumptions
}\label{sec:assumptions}

In this section we take advantage of the speed of our rapid binary
evolutionary code and population synthesis routines, which enables us to evaluate the impact of our assumptions.  
This concerns the initial distribution functions (Sect.~\ref{uncertainDist}) and assumptions about the physics of binary evolution (Sect.~\ref{uncertainMod}).
We quantify the effects by comparing the fraction of stars with projected rotational velocities in excess of 200\kms, $F_r$, that results from the various assumptions. We use our standard simulation for $L>10^4\Lsun$ as reference.

\begin{table}
\caption{Impact of the assumed initial binary fraction and the initial distribution of binary parameters (as defined in Sect.~\ref{pop}) on the fraction $F_r$ of stars with $\vsini>200\kms$ \label{tab1}.  For discussion see Sect.~\ref{uncertainDist}. }
\begin{center}
\begin{tabular}{lcc}
\tableline\tableline
initial distribution  & variation & $F_r (\%)$ \\
\tableline
standard$^*$  &&  18.8 \\ 
orbital period &    $0 >\pi>-1   $              &  13.1 -- 24.7 \\
mass ratio &   $-1 <\kappa<1 $               &   14.8 -- 21.3 \\  
primary mass &  $1.65 <\alpha< 3.05$   &  17.0 -- 21.3\\ 
binary fraction &   $0.5 <\fbin<1.0$        &  14.6 -- 23.9 \\
\tableline
\end{tabular}
\tablecomments{ $^*\pi =-0.5$, $\kappa=0.0$, $\alpha=2.35$ and $\fbin = 0.7$}
\end{center}
\end{table}

\subsection {Initial distribution functions}\label{sec:assumptions-distributions}
\label{uncertainDist}

The effect of uncertainties in the adopted initial distributions is relatively small, each affecting  the fraction of rapid rotators only by a few percent in absolute terms, see  Table~\ref{tab1}.  The \emph{initial orbital period} distribution  is the largest cause of uncertainty accounting for a variation of $\pm5.8\%$. When we compare a distribution that is flat in $\log P$, i.e \"Opik's law ($\pi = 0$) to a distribution that strongly favors short period systems ($\pi = -1$), we find that  $F_r$ increases  from 13.1 to 24.7\%.  The latter distribution favors systems that  interact while the primary star is still on the main sequence. In our simulations, these systems produce rapid rotators through mass transfer and mergers that are on average more massive, and thus brighter, than rapid rotators produced in wider binaries.    

Uncertainties in the adopted \emph{distribution of mass ratios} are responsible for a variation of just over $\pm3\%$  when we vary the power law exponent from $\kappa=-1$ to $\kappa =1$, which generously allows for the uncertainties quoted by \citet{Sana+2012}.   A distribution skewed towards systems with comparable masses for the components ($\kappa =1$) favors systems that produce rapid rotators through nearly conservative mass transfer. The spun up mass gainers are typically brighter than the original primary and they 
experience significant rejuvenation.  Both effects result in a larger fraction of rapid rotators in the population.  A distribution that favors systems with extreme mass ratios ($\kappa = -1$) results in a larger number of mergers.   While mergers in short period systems contribute to the fraction of rapid rotators, mergers resulting from wider systems do not as we consider these to become post main-sequence objects,  explaining the reduction of the fraction of rapid rotators  
 
The rapid rotators in our simulation result from stars that gain mass, apart from the small contribution of tidally spun up stars. On average a gain in mass implies a larger luminosity.   Stars that were initially not bright enough to be included in our distribution can become sufficiently bright after accreting.  Changing the relative number of lower mass systems, by adapting the \emph{initial mass function},  increases the number of potential systems that can produce rapid rotators. However, the effect is small: varying the steepness of the initial mass function of the primary star by $\alpha = 2.35\pm 0.7$ leads to a change of about $\pm2\%$.

The effect of varying the initial binary fraction can be obtained by rescaling the results. We account for the fact that the fraction of stars born as single stars in the population $F_s$ is not equal to the fraction of single stars at birth. The fraction of rapid rotators $F_r$ for any given initial binary fraction \fbin~ is given by
\begin{equation}
F_r (\fbin) = F_r^* \frac{ \fbin }{\fbin^*}  \left(   F_b^* \frac{ \fbin }{\fbin^*} +  F_s^*  \frac{1-\fbin }{1-\fbin^*}    \right)^{-1},
\end{equation}
where superscript $^*$ refers to the values obtained in our standard simulation (see Fig.~\ref{vrot_dist})  and $F_b = 1-F_s$.  Considering a population in which only half or all stars are born as binaries with mass ratios and orbital periods as specified in Sect~\ref{pop} results in variations in $F_r$ of about $\pm 5\%$.

\subsection {Uncertain physics}
\label{sec:assumptions-physics}
\label{uncertainMod}
In this section we address uncertain physical assumptions regarding the treatment of  mass transfer, the formation of contact systems and mergers.

\subsubsection {Do contact systems merge? \label{contact}}
Mass accretion drives the companion star out of thermal equilibrium causing it to expand when the mass transfer timescale is short compared to the thermal timescale of the companion \citep[e.g.][]{Neo+1977}.  This can lead to the formation of a contact binary \citep{Pols1994a, Wellstein+2001}.   However,  the parameter range for which contact occurs is uncertain as it depends, among other things, on the poorly constrained specific entropy of the accreted material \citep{Shu+1981}.  

When contact is established, it is not clear whether this always leads to a common envelope phase and a merger. If the secondary star overfills its Roche-lobe by only a small amount,  the star may detach again as the star restores thermal equilibrium \citep{de-Mink+2007}.  However, the outcome depends on mass and angular momentum loss during this phase.  If the stars come into deep contact, mass loss from the system through the outer Lagrangian point efficiently drains angular momentum from the system and drives the stars deeper into contact. Mass loss through stellar winds or bipolar outflows can, in principle, have the opposite effect. 

In our standard simulation we account for the formation of contact due to rapid mass transfer by considering the mass ratio $q = M_2/M_1$ at the onset of Roche-lobe overflow.  If the mass ratio is smaller than a critical value we assume that contact is established and results in a merger.  If the mass ratio is larger, we continue to compute the accretion of mass and angular momentum onto the secondary star, and assume that contact is avoided as long as the thermal equilibrium radius of the mass gainer is smaller than its Roche radius.   Below we discuss the impact on our results of varying the adopted critical mass ratio systems that interact during the main sequence $q_{\rm crit, MS}$ and systems in which the donor star is a post main-sequence star crossing the Hertzsprung gap $q_{\rm crit, HG}$.

\paragraph {Contact involving two main-sequence stars}  
 Systems that interact during the main sequence eventually produce a rapid rotator either through a merger or through spinup by during mass transfer regardless of their initial mass ratio. Therefore the fraction of rapid rotators is not very sensitive to the adopted critical mass ratio for merging $q_{\rm crit, MS}$. In our standard simulation we adopted $q_{\rm crit, MS} = 0.65$.   If we vary $q_{\rm crit, MS}$ in the range  0.8--0.25, the fraction of rapid rotators changes by only $\pm 0.5\%$. 
 On the one hand, mergers are typically more massive and more luminous than the secondary star would have been had the system not merged.   As a result, in our brightness limited sample the rapidly rotating stars resulting from mergers originate from binary systems with a wide range of initial masses and mass ratios.  On the other hand, we find that the remaining relative lifetime of mergers is typically much smaller than that of rapid rotators produced through mass transfer. Mergers contain the helium rich core of the original primary and even though some fresh hydrogen is mixed into the central regions, their remaining lifetime is limited.  In contrast, the core of the secondary is still hydrogen rich at the onset of mass transfer in comparison to the core of primary.  In the population, the longer remaining lifetime of the secondaries compensates for their lower brightness.  
 
\paragraph {Contact involving Hertzsprung gap donor}  For post main-sequence interaction the formation of contact introduces a larger uncertainty on the predicted number of rapid rotators.  If a binary system mergers after the primary star has left the main sequence, the product is expected to become a post main sequence object, a blue or red super giant. Hence, we do not include such stars in our distribution of rotation rates.  In our standard simulation we assume that such systems merge if the mass ratio at the onset of Roche lobe overflow is smaller than $q_{\rm crit, HG} = 0.4$.   Detailed binary evolutionary models show that the parameter range for the formation of contact may be substantially larger \citep[e.g.][]{Wellstein+2001}. For the extreme assumption that all these systems merge we find an absolute change of $-5.4\%$ for the fraction of rapid rotators.  To obtain this estimate, we mimic the results by \citet{Wellstein+2001} by assuming that all systems with initial orbital periods larger than 30 days and mass ratios smaller than 0.65 merge.  Even though this assumption affects a very large fraction of the initial parameter space, the effect is limited because it mostly affects systems that would produce relatively low mass rapid rotators.  Taking the opposite extreme assumption that none of the Case B systems merge ($q_{\rm crit, HG} = 0.0$) does have a significant effect on the fraction of rapid rotators, for the same reasons.

\subsubsection {Can stars continue to accrete after they are spun up to Keplerian rotation? \label{beta}} 
 \citet{Packet1981} argued that a star fed by an accretion disk reaches Keplerian rotation after accreting just a few percent of its initial mass.   It remains unclear whether a star rotating at this speed  can continue to accrete mass.  In the models by \citet{Petrovic+2005} and \citet{de-Mink+2009} it was assumed that accretion ceases when the star reaches the Keplerian limit.  Only in short period systems tides can spin the accreting star down such that it can undoubtedly resume accreting a substantial amount of mass.    \citet{Popham+1991} argue that viscous coupling between the star and the disk can govern an  inward flow of mass and an outward flow of angular momentum such that also mass gainers in wider binaries may continue to accrete while rotating near or at the Keplerian limit.  While we are not in a position to resolve this debate, we have the tools to investigate the impact of both extreme assumptions on the distribution of rotation rates.  In our standard simulation we followe \citet{Popham+1991}.  Following \citet{Petrovic+2005} instead does not reduce the  number of stars that are spun up, but it does reduce the mass and therefore brightness of these stars.  As a result we find that the fraction of rapid rotators in our brightness limited sample changes by $-11\%$.   

\subsubsection  {Mixing and mass loss during mergers?}
Mergers constitute the least understood phase of binary evolution. In particular, the amount of mass loss and the amount of mixing is uncertain.  In our standard simulations we assume that a fraction $\mu_{\rm loss} = 0.1$ of the total system mass is lost during a merger and that a fraction $\mu_{\rm mix}=0.1$  of the envelope is mixed into the core.   Changing the amount of mass loss affects the number of mergers that are bright enough to be included in the sample, but the effect is small.   Varying  $\mu_{\rm loss}$ in the range 0--0.25 results in a change that is not larger than $\pm 0.4\%$.    Changing the amount of mixing primarily changes the remaining lifetime of the merger product.   Varying $\mu_{\rm mix}$ in the range 0--1 to simulate the effect of no additional mixing versus complete mixing of the core and the envelope we find a change of  $-0.8\%$ and $+2.5\%$,  respectively.

\subsubsection{Magnetic fields?}
\label{Bfields}
Recent searches have revealed large scale magnetic fields in several O type stars \citep[e.g.][]{Wade+2012}  with strengths large enough to affect their spin down time \citep{Ud-Doula+2009}.  Several recent findings may be relevant for our study. The apparent single O~star HD\,148937 has shown to host a large-scale magnetic field \citep{Hubrig+2008, Hubrig+2011, Naze+2010, Wade+2012a}, and is a strong candidate for being a merger product \citep{Langer2012}.  It is surrounded by a bipolar nebula that may have been ejected during the merger event.   Furthermore, the recent detection of a large scale magnetic field in the mass gainer of Plaskett's star \citep{Grunhut+2012} implies that such fields can exist in a star shortly after an accretion event, and may even indicate that such fields can be generated as its result.  In this context it is interesting to mention the tentative detection of a magnetic field in the O star of Cyg-X1 by \citet{Karitskaya+2010}, see however \citet{Bagnulo+2012}, and for HD\,153919 \citep{Hubrig+2011}, which is the O star companion to the  $\sim2.5\Msun$~compact object 4U1700-37 \citep[cf.][]{Clark+2002}.

If the production of a large-scale magnetic field would be a common by-product of a mass transfer or merger event \citep[e.g.][]{Ferrario+2009, Tutukov+2010} or \changes{if could be generated as a result of a rotationally driven dynamo  \citep[e.g.][]{Potter+2012}}, we would have underestimated the spin down times of such products \citep[e.g.][]{Meynet+2011} . Since this scenario is speculative at the moment, we refrain from quantifying such an effect, but it might reduce the number of rapid rotating stars produced by binary interaction. It would not, however, decrease the number of binary products in a given population \citep{de-Mink+2012-letter}.

\begin{table*}
\caption{ A compilation of observational studies listing the percentages of stars, $F$, with projected rotational velocities larger than 200, 300, and $400\kms$.   \label{compilation}}
\begin{center}
\begin{tabular}{ l l  c c c c c c}
\tableline\tableline

\hspace{.3cm}Reference                                   & Environment   & Sp.Type & Sample size  & $F_{>200\kms}$ & $F_{>300\kms}$ & $F_{>400\kms}$ \\
\hspace{.3cm}                               &    &  &   & (\%) &  (\%) &  (\%) \\
\tableline
\\
\hspace{.0cm}\emph{Observed samples}  \\
%\hspace{.3cm}\citet{Slettebak1949}    & MWG          &  O       &  21 &      33      &    19       &    10        \\
\hspace{.3cm}\citet{Conti+1977}                            & MW          &  O       & 205 &      26      &    10       &     0        \\
\hspace{.3cm}\citet{Howarth+1997}                        & MW          &  OB      & 373 &       9      &     4       &     1        \\

\hspace{.3cm}\citet{Abt+2002}                            & MW          &  B0-3    & 357 &      16      &     4       &     0.3      \\

\hspace{.3cm}\citet{Martayan+2006}                       & LMC (NGC~2004) &  B       & 121 &      15      &      7      &     0        \\

\hspace{.3cm}\citet{Martayan+2007}                      & SMC (NGC~330)   &  B       & 198 &      27      &      9      &     1.0      \\

\hspace{.3cm}\citet{Hunter+2008}                         & LMC          &  OB      & 204 &      16      &     2       &     0        \\
\hspace{1cm}"\hspace{1cm}"                                           & SMC          &  OB      & 204 &      22      &     6       &     0        \\

\hspace{.3cm}\citet{Penny+2009}                          & MW          &  OB      &  97 &      22      &    10       &     1        \\
\hspace{1cm}"\hspace{1cm}"                                            & LMC          &  OB      & 106 &       8      &     9       &     0        \\
\hspace{1cm}"\hspace{1cm}"                                            & SMC          &  OB      &  55 &      13      &     5       &     0.2      \\
\hspace{.3cm}\citet{Huang+2010}                          & MW (Clusters) &  B0-9    & 695 &      28      &     7       &     0.1      \\
\hspace{1cm}"\hspace{1cm}"                                          & MW (Field)    &  B0-9    & 483 &      19      &     5       &     0        \\

\emph{VLT-FLAMES Tarantula Survey} (VFTS)   &&&&&& \\
\hspace{.3cm}Dufton et al. (2012, subm.)   & LMC (30~Dor)  &  O9.5-B3 & 337 &      50      &    15       &     3        \\
\hspace{.3cm}Ram\'{\i}rez-Agudelo et al.  (in prep.)   & LMC (30~Dor)    &  O       & 178 &      30      &    16       &     5        \\

\tableline
&&&&&& \\
\hspace{.0cm}\emph{Predictions from this work}  \\
\hspace{1cm}"\hspace{1cm}"                      &     all  &    &&     19       &    11        &     2        \\
\hspace{1cm}"\hspace{1cm}"                         & mergers  &    &&      5       &     3        &     0.5      \\
\hspace{1cm}"\hspace{1cm}"                         &   other  &    &&     14       &     8        &     1.5      \\
\tableline

\end{tabular} 
\end{center}
\end{table*}

\changes{

\section {Discussion and implications}

 Based on our simulations we expect $20_{-10}^{+5}$\% of massive main-sequence
stars to rotate rapidly as a result of binary interaction. In this section we
discuss implications for rotationally
induced mixing in stars (Sect.~\ref{disc_rot}), and for our understanding of
the origin of Be stars (Sect.~\ref{disc_Be}).  We place our results in
context of the observed distribution of rotation rates (Sect.~\ref{disc_dist}), 
and we discuss how to proceed from here to derive the true birth spin
distribution of massive stars (Sect.~\ref{disc_birth}).

\subsection{Implications for rotationally induced mixing
\label{disc_rot}}

Rotation has been argued to be a major factor influencing the evolution of
massive stars. In particular, it is thought to induce mixing in the radiative
layers of massive stars \citep[][and references therein]{Maeder+2000a,
Heger+2000}. The presence of hydrogen-burning products at the surface of O and B
type stars has generally been interpreted as a signature of rotational mixing
\citep{Gies+1992,Hunter+2008a,Przybilla+2010}.  \citet{Hunter+2008a} analyzed
about one hundred early B-type main-sequence stars in the LMC with projected
rotational velocities of up to $\sim$300\kms.  The rapid rotators with enhanced
nitrogen abundances in this sample have been considered to provide the most direct
evidence for rotational mixing \citep{Maeder+2009,Brott+2011}.

Our results question the validity of this interpretation; they suggest that a
significant fraction, or perhaps even close to all of the rapid rotators are the products of
close binary evolution \citep[see][for a potential exception]{Acke+2008}. 
Detailed models show that mass transfer in a binary
results in surface nitrogen enrichment that covers the observed range
\citep{Wellstein+2001, Langer2012}. If the observed nitrogen enrichment of the
rapid rotators is primarily the result of mass transfer, 
there must be less room for rotational mixing \citep{Langer+2008, Brott+2011a}.

Additional evidence for the importance of close binary evolution in the sample
of \citet{Hunter+2008a} comes from the star-by-star analysis of \citet{Kohler+2012}. They conclude 
that 10 of the rapidly rotating stars in the sample can not be explained by 
rotating single stellar models; the observed nitrogen abundance for these stars
is much lower than expected for their age and their projected rotation rate.
To reconcile these objects with the theory of rotational mixing 
requires to assume that these stars were slow rotators for most of their lives,
and that they have been spun-up only recently due to close binary interaction by
non-conservative mass transfer.
We conclude that it is (so far) difficult to single out the effect of rotational mixing in samples
of massive main sequence stars. It turns out that
the strategy to circumvent this problem by removing identified binaries from
the observed sample is not effective; it may even achieve the opposite.  As 
argued in \citet{de-Mink+2011}, the products of binary evolution typically are or appear
to be single stars. 
As most of the stars which are found to be
a member of a binary system did not yet gain any mass from their companion,
removing stars with evidence for binarity from the  
sample may in fact increase the relative contamination of the sample with 
post-interaction binary products
\citep{de-Mink+2012-letter}.

We conclude that the role of rotational mixing remains ambiguous at present.
This is unsatisfying given the large implications of rotational mixing on the
evolution of the progenitors of long gamma-ray bursts and pair-instability
supernovae \citep[e.g.][]{Yoon+2006, Woosley+2006, Langer+2007} as well as on   
the chemical and radiative feedback of massive stars, in particular in the
early Universe \citep[e.g.][]{Ekstrom+2008a, Yoon+2012}. 
Further studies of the relation between rotation rates and nitrogen surface abundances of
large and well-defined samples of stars are essential. A first step may be to establish whether 
the types of behavior seen in B stars are also present in O stars.

\subsection{ Evidence for spin-up from Be/X-ray binaries and the binary origin  
of Be stars \label{disc_Be}} 

Compelling evidence for spin-up of massive
main-sequence stars as a result of binary interaction comes from the Be/X-ray binaries,  
the most frequent type of high mass X-ray binaries \citep{Liu+2006}. This class, first
recognized by \citet{Maraschi+1976}, consists of a neutron star in an eccentric orbit
around a rapidly rotating B-main sequence star. The formation of these systems can be
understood by assuming that the progenitor of the neutron star transferred mass
and angular momentum to its companion before it exploded as a
supernova \citep{Rappaport+1982}. The Be/X-ray binaries constitute the subset of
systems that remained bound after the explosion.  The majority of the systems   
is, however, expected to be disrupted as a result of the kick of the neutron  
star, thereby producing single Be stars \citep[e.g.][]{Blaauw1961, Eldridge+2011}.

The idea that a significant fraction, or even all, of the Be stars result from
binary interaction has been explored by \citet{Pols+1991} and
\citet{van-Bever+1997}.  Both studies conclude that binaries provide a significant 
fraction of single Be stars, but they also find that the number of mass transfer 
systems is not high enough to explain all Be stars. However, our models  show 
that binaries produce significantly more rapid   
rotators than quoted in these studies. This is partially due to the fact that we
consider stellar mergers as a channel to form rapid rotators.  Furthermore, the 
initial binary parameter distributions adopted in these original studies were based on
\citet{Abt1983} and \citet{Abt+1978}.  We adopt distributions based on
recent work \citep{Sana+2012}, which made clear that earlier studies
underestimate the frequency of short period binaries for O-stars
\citep{Sana+2012}

\changes{We refrain from predicting the fraction of Be stars 
produced by our models because the precise
conditions for the Be-phenomenon remain somewhat elusive.  It is however worth  
to mention that the total fraction of mergers and mass gainers in our standard
simulation is 24.1\%, (cf.\ the top right panel of Fig.4). For comparison, the
fraction of Be stars among early-type, non-supergiant B-type stars is 20-30\%
after correcting for selection effects \citep{Zorec+1997}. }
  
\subsection{Comparison with observed samples \label{disc_dist}}

A direct comparison between the various observational results, summarized in
Table~\ref{compilation}, and our predictions is difficult, due to different
selection and methodological effects, and potentially different age
distributions of the respective stellar populations. For example, some
observational studies attempt to identify and remove spectroscopic binaries
while others did not. Furthermore, some samples are biased against rapid
rotators, for example because emission line stars were not included or because
of an inherited bias in samples that are based on archival spectra.  A
comparison with our predictions should thus also be taken with care. Overall,
however, we find remarkable trends which appear to be present in all samples.

The fraction of stars with $\vsini > 200\kms$ in the observed samples compiled
in Table~\ref{compilation} ranges between 10 and 50\%, while our calculations
predict numbers of the order of 20\%, all produced from binaries. Our
theoretical number is based on the assumption of a constant star formation rate.
We expect that for certain star formation histories, e.g. a starburst, and
certain observational biases, e.g. in samples focusing on the cluster turn-off,
the rates may be significantly different over the situation captured by our  
models \citep[e.g.][Schneider et al. 2012, in prep]{Pols+1994, van-Bever+1998,
Chen+2008}.

Considering higher cut-off values for \vsini\ appears to leave less and less
room for single stars, even considering the error margin in our prediction (cf.,
Sect.~5).  Above 300\kms, only the Dufton et al. (2012, subm.) and
Ramirez-Agudelo et al. (2013, in prep.) samples, both resulting from the
VLT-FLAMES Tarantula Survey (VFTS), find more stars than we predict. The latter
works however have removed detected spectroscopic binaries from their sample,
which account for almost a third of their total sample. As can be seen from   
Fig.~4, spectroscopic binaries are expected to mostly contribute to the
\vsini~distribution below $300\kms$.  If we correct our predictions for this we
obtain $F_{>300\kms} = 16\%$, which is remarkably close to what is observed in 
the VFTS samples.

This trend continues when considering stars above 400\kms. While the number of  
observed stars is small, the associated percentages from the VFTS samples are close to our results.
We note that the VFTS samples do not suffer from an explicit bias against the
most rapid rotators, in contrast to the other samples.  Whether this is
sufficient to explain the observational differences with the VFTS results or
whether there is a genuine difference between the rotational properties of stars
in 30 Dor and stars in less extreme environments would require a careful
assessment of the selection effects and observational biases of these other
samples. 

The above analysis suggests a decreasing fraction of genuine single stars in
stellar samples with increasing \vsini~cut-off. It appears possible that there 
is a \vsini~ threshold above which binary products dominate the population. To
establish such a threshold, and its error bar, requires tailored binary
population models which reproduce specific observed samples; this is beyond the
scope of this paper. Nevertheless, from the preliminary comparison of our models
with the existing data in Table~\ref{compilation} it seems conceivable that this
threshold value --- which will depend on metallicity, star formation history and
sample biases --- could generally be smaller than 300\kms.

\subsection{Towards deriving the initial distribution of rotation rates   
\label{disc_birth}}

Deriving the initial distribution of rotation rates is a high priority
given its importance as a constraint for star-formation theories
\citep[e.g.][]{Rosen+2012} and as input condition for modeling populations of  
massive stars both nearby and at high redshift \citep{Brott+2011a,
Levesque+2012, Eldridge+2009}.  In the literature, it is often assumed that the distribution of
rotation rates of early-type stars closely reflects their birth spin
distribution, arguing that the rotational velocity of massive stars is not
expected to change significantly during their main-sequence evolution --- which
we do confirm when considering only genuine single stars. However, we have shown
that this assumption does not generally hold,
due to the large close binary fraction of massive stars \citep{Sana+2012} and   
the consequences of close binary evolution.

For star clusters or a star-formation rate which is strongly peaked in time, the
effects may be smaller (e.g. for extremely young star clusters) or larger
(looking at the main-sequence turn-off of star clusters, where all stars are
close to ending core hydrogen burning). Since star clusters that are much
younger than the lifetime of their most massive stars are difficult to find, it
may remain a challenge to derive the true birth distribution of the spins of
massive stars directly from observations.

There may be two promising ways to proceed. One may try to observe the birth
spin distribution of massive stars directly in young star clusters, excluding
the most massive stars.  If such kind of studies are successful, the \vsini-distribution can be compared with
those of stars in the same mass range, for older clusters, where binary effects
had time to operate. This could be compared with results from our
method, with tailored input functions appropriate for the observed samples,
which would allow to draw conclusions about the effects of binary evolution.
Alternatively, our method may be used in the future to directly derive the
initial spin distribution of an observed population, adopting an iterative
process where this distribution is varied until the observations are
appropriately reproduced.
}

\acknowledgments{{\it Acknowledgements:} \footnotesize SdM acknowledges the Argelander Institute of Bonn University for hospitality during various visits.  We thank sthe members of the VLT-FLAMES massive stars consortium (PI C. Evans) and especially Ines Brott, Philip Dufton,  Paul Dunstall, John Eldridge, Evert Glebbeek, Eveline Helder, Danny Lennon, Colin Norman, Stan Owocki, Oscar Ram\'irez-Agudelo, Fabian Schneider and Nolan Walborn, for fruitful discussion or commenting on early versions of this paper.  In particular, we are indebted to Onno Pols for comments and his contributions and updates made over the years to the original fortran version of the code.  Furthermore we thank the referee for suggestions to enhance the clarity of the discussion. 

Support for this work was provided by NASA through Hubble Fellowship grant HST-HF-51270.01-A awarded by the Space Telescope Science Institute, which is operated by the Association of Universities for Research in Astronomy, Inc., for NASA, under contract NAS 5- 26555. We also acknowledge partial support by the National Science Foundation under Grant No. 1066293 and the hospitality of the Aspen Center for Physics. 
}

\appendix

\section{Rapid binary evolutionary code}

In Section~\ref{code} we briefly described the rapid binary evolutionary code employed in this study.  Here, we provide a more complete description with references and document routines that have not been documented elsewhere. 

%%%%%%%%%%%%%%%%%%
\subsection {Stellar wind mass loss}
%%%%%%%%%%%%%%%%%%
To account for the radiatively driven winds of massive stars we
implement the mass-loss prescription of \cite{Vink+2000, Vink+2001}.
This recipe accounts for the fast increase of the mass-loss rate for
stars with temperatures below $22,\!000\,\mathrm{K}$, which is related to the
recombination of Fe IV to Fe III and is commonly referred to as the
bi-stability jump. Following \citet{Brott+2011}, we perform a linear interpolation in the
mass loss rate within $2,\!500 \mathrm{K}$ of the jump, to
ensure a continuous transition.

To accommodate the strong mass-loss increase when
approaching the \citet{Humphreys+1994} limit, we switch to the
empirical mass loss rate of \citet{Nieuwenhuijzen+1990} at any
temperature lower than the critical temperature for the bi-stability
jump, when the \citet{Vink+2001} rate becomes smaller than that
from \citet{Nieuwenhuijzen+1990}. This ensures a smooth transition between the two mass loss
prescriptions and naturally accounts for the increased mass loss at
the theoretically predicted second bi-stability jump at $\sim 12,\!500 \mathrm{K}$.

We adopt Wolf-Rayet mass-loss rates of \citet{Hamann+1995} for naked
helium stars and stars with thin hydrogen envelopes \citep{Hurley+2000}.  To account for the
extreme mass loss of stars beyond the Humphreys-Davidson limit
limit an LBV-like mass loss rate is included as described in
\citet{Hurley+2000}. Mass-loss rates during the evolved stages of
intermediate and low-mass stars are based on \citet{Kudritzki+1978}
and \citet{Vassiliadis+1993}, see \citet{Hurley+2000}.

% \paragraph {Metallicity dependence}
The prescription by \citet{Vink+2001} accounts the metallicity
dependence of radiatively driven winds with a scaling factor of
$(Z/0.02)^{0.85}$, where $Z$ is the metallicity expressed as initial
mass fraction of elements heavier then helium. We apply the same
scaling for the \citet{Nieuwenhuijzen+1990} rates. Other rates
are assumed not to depend significantly on metallicity
\citep[e.g.][]{van-Loon2006}.

%%%%%%%%%%%%%%%%
\subsection {Effects of rotation}
%%%%%%%%%%%%%%%%
Throughout the evolution of each binary system we
compute how the stellar spins change as a result of their internal
evolution and angular momentum loss and gain by tides, mass exchange
and stellar winds.

\subsubsection {Internal evolution and the moment of inertia}

As the star evolves its moment of
inertia, $I = k M R^2$ changes, where $k$ is the radius of gyration
squared.  The original implementation described in \citet{Hurley+2000}
assumed constant $k=0.1$ during the main sequence.  However, even over the course of the
main sequence the evolution of the internal structure significantly changes \citep[e.g.][]{Ekstrom+2008}.
Therefore, we adopt fitting formula (Pols, priv. communication) for
the evolution and mass dependence of the gyration radius based on evolutionary models by \citet{Pols+1998}.   For zero-age main-sequence stars the gyration radius squared, $k_0$, is given by
\begin{equation}
    k_{0} \simeq  c +  \min \{ 0.21,\,\,   \max  \{     0.09 - 0.27 \log M,\,\, 
     0.037 + 0.033 \log M \} \},
\end{equation}
where the correction factor $c=0$ except for stars with 
$\log M >1.3$, for which $c = -0.055 (\log M -1.3)^2$.
When the stars evolve along the main sequence, the star becomes more
centrally condensed and the gyration radius decreases. This can be
described in terms of the radius $R$ and the radius at zero-age $R_0$,
\begin {equation}
k \simeq (k_{0} - 0.025) \left(\frac{R}{R_{0}}\right)^C +
0.025\left(\frac{R}{R_{0}}\right)^{-0.1},
\end{equation}
where
\begin{equation}
  C  = \left\{
    \begin{array}{ccc}
      -2.5  & \mbox{} & \log M < 0,  \\
      -2.5 + 5 \log M & \mbox{for} & 0 < \log M <0.2,   \\
      -1.5  & \mbox{ } & 0.2 < \log M.  
    \end{array}  \right. 
\end{equation}

Although our fits are based on models at a metallicity $Z=0.02$, we account for the main effect of metallicity through the dependence of the radii on metallicity,  as a result of the choice to express $k$ in terms of  $R/R_0$.  This approximation is sufficient for our purposes.  Analytic approximations of
the evolution of the gyration radius at more evolved evolutionary stages are included as well, but the differences with respect to the original implementation by \citet{Hurley+2000} do \changes {not} significantly affect our study.  

We assume that the internal rotational profile can be approximated by the assumption of rigid rotation as a result of efficient internal transport of angular momentum.   For main-sequence stars this is supported by the findings of detailed stellar models \citep{Ekstrom+2008, Brott+2011} independent of the detailed treatment of the internal angular momentum transport processes.

\subsubsection {Deformation by rotation and the Keplerian limit }

Stars become oblate when they rotate near their Keplerian limit, i.e. the rotation rate for which material at the equator is no longer bound to the star because the outward centrifugal acceleration balances the inward gravitational acceleration. Deviations from spherical symmetry have been observed
directly by interferometric studies, for example Altair
and Alchernar \citep{Peterson+2006, Carciofi+2008}.
We account for deformation of rotation as described in Sect.~\ref{code}.  To obtain the equatorial radiii 
we compute the shape of the equipotential surface
in the Roche approximation. Under the assumption that the polar radius
is not affected by rotation this becomes an algebraic equation of the
third degree \citep[e.g.][]{Maeder2009}.  We find that the numerical
solution obtained by the Newton method can be approximated by the
following analytic approximation  
\begin{equation}
\log_{10}  \frac{R_{\rm eq}}{ R_{\rm p}} (\omega) 
  \simeq   \omega \, \left( f_1  \tan [1.422 \,\omega]   
      +   f_2 \sin[\omega] \right),
\end{equation}
where $f_1 = 1.7539\times 10^{-2} $ and $f_2 = 4.511\times10^{-2}$, which can be evaluated in the timely fashion required for population synthesis. We note that there is also some ambiguity in the literature with respect to these two possible definitions of `the fraction of break-up' expressed in terms of the the rotation rate, $\Omega$ and in terms of the rotational velocity $v_{\rm eq}$. Note that $\Omega / \Omega_{\rm K} \ne v_{\rm eq} / v_{\rm eq, K}$ \citep[e.g.][]{Maeder2009}.  For slow and intermediate rotation rates the fraction of break-up in terms of the rotation rate and  the fraction of break-up in terms of the rotational velocity deviate by a factor 0.67  since 
\begin{equation}
  \frac{v_{\rm eq}}{v_{\rm eq, K}}  = \frac {R_{\rm eq}}{R_{\rm eq, K}}  \Omega / \Omega_{\rm K}.
\end{equation}

\subsubsection { The interplay of rotation and the stellar wind}
To account for the effect of rotation on the mass loss by stellar
winds we follow the approach by
\citet[][Eq.~4.30]{Maeder+2000}.  They derive the latitudinal mass
flux from a star deformed by rotation taking into account the effect
of gravity darkening \citep{von-Zeipel1924},
\begin{equation}
\frac{\dot{M} (\omega) }{ \dot{M} (0)} \simeq \left(\frac{ 1-\Gamma }{
    1 - \Gamma - f(\omega) }\right)^{\frac{1}{\alpha} -1},
\end{equation}
where $\Gamma$ is the Eddington factor and $\alpha$ the force
multiplier, for which we use the empirical relation by \citep{Lamers+1995} valid for $\log T_{\rm eff}=
  3.9-4.7$, where $\alpha$ varies from 0.15-0.52. Outside this range
  we assume that $\alpha$ is independent of the effective
  temperature. The dependence on the rotation rotation rates enters
through
\begin{equation}
f(\omega) = \frac{\Omega}{2 \pi G \rho_m}
% \approx \frac{4}{9}\left[ \omega \frac{R_{\rm eq}(\omega)}{R_{\rm
%   eq, K}} \right]^2
\approx 0.198\left( \omega \frac{R_{\rm eq}}{R_{\rm p}} (\omega)
\right)^2.
\end{equation}
The approximation on the right hand side is accurate up to $\omega
\lesssim 0.8$.
We compute the Eddington factor,
\begin{equation}
\Gamma = \frac{\kappa L}{4 \pi c G M}
\end{equation}
by approximating the opacity $\kappa$ with the electron scattering
opacity, $\kappa \simeq 0.2(1+X)$, where $X$ is the mass fraction of
hydrogen at the surface. Since we do not explicitly follow the surface
composition we adopt $X=0.74$ for stars that still have a hydrogen
envelope and $X=0$ for naked helium stars.

In this prescription the effect of rotation on the mass loss rate is small \citep{Maeder+2000}. Only for $\Gamma > 0.64$ and $\log T_{\rm
  eff} \leq 4.3$  is a large increase of the mass loss rate
predicted. In all cases we limit the stellar wind mass loss rate to the thermal mass loss rate $\dot{M}_{\rm KH} \equiv    M / \tau_{\rm KH}$, where the we approximate the Kelvin Helmholtz timescale $\tau_{\rm KH}$ as  
\begin{equation}
 \tau_{\rm KH}  = 10^7\,\mathrm{yr}  \frac{M} {R L},  
\end{equation}
where M, R and L refer to the current mass, radius and luminosity expressed in solar units. 

The radiatively-driven stellar wind of a star deformed by rotation becomes aspherical as a result of the changes of the effective gravity with latitude.   \cite{Georgy+2011} computed the net effect of the latitudinal dependence on the loss of angular momentum. They find that the deviation from the spherical cases is very small. In the interest of computational speed we can therefore safely assume that the specific angular momentum of mass lost through the stellar wind is equal to $j = 2/3 R_p^2 \Omega$.

\subsubsection {Mass and angular momentum loss of stars rotating close to Keplerian rotation}
During their life stars may approach the Keplerian rotation rate as a result of mass accretion
\citet{Packet1981} or as a result of internal evolution. \citet{Ekstrom+2008}.
In this case the star must dispose of its excess angular momentum at a rate
dictated by the process that drives the star to its Keplerian limit.
If angular momentum loss by the stellar wind or tides is not efficient enough, it is likely that a near-Keplerian disk is formed
 when the thermal motion of gas particles at the equator becomes sufficient to overcome the escape speed, such that they are launched in orbit around the star \citep{Okazaki+2001}. Viscous coupling of the disk and the star may
help extract angular momentum efficiently \changes{\citep{Paczynski1991, Sills+2005, Krticka+2011}}, while
losing only very little mass, depending on the viscosity and the
extent of the disk.  The disk may be truncated by the torque of the
companion star.  A further possibility is the illumination of the disk
which may cause it to flare and disperse. The interplay between these processes will determine how much mass the rapidly rotating star can accrete.  A detailed treatment of outflowing disks within a binary system is beyond the purpose of this study, but we discuss the effects of extreme assumptions in Section~\ref{sec:assumptions-physics}.

\subsection{Treatment of binary interaction}

\subsubsection {Mass transfer rate}
We model mass transfer as described in \citet{Hurley+2002}, but with the
following adaptations.  In the original version of the code the
rate for stable mass transfer is assumed to be a steep function of the
amount by which the star overfills its Roche lobe
\citep[][eq.~58]{Hurley+2002}.  This prescription can introduce
numerical instabilities causing the star to oscillate in and
out of its Roche lobe.
Therefore, we replace this prescription by an adaptive algorithm.
When the star fills its Roche lobe we determine the mass transfer rate
by removing a small amount of mass and computing the
resulting change in radius until the stellar radius becomes smaller
than the Roche lobe radius within a certain threshold.  The rate is
capped by the thermal rate as defined in \citet[eq.~60 ]{Hurley+2002}.
As a result, the mass transfer rate is governed by the
evolutionary changes in the stellar radius and changes in the
Roche lobe.

When a star accretes mass on timescales that are short compared to the
thermal time scale the accreting star may be
driven out of thermal equilibrium resulting in expansion
\citep{Benson1970a}.  How severe this effect is is uncertain as it depends on the poorly constrained specific entropy of the accreted material. Shocks
during the accretion process may result in the accretion of lower
entropy material \citep{Shu+1981}. Most detailed binary evolutionary
codes use a simple boundary conditions where material is assumed to be
accreted with the specific entropy of the surface of the accretor.
This leads to severe expansion in binaries with extreme mass ratios implying that a large fraction of the systems will
come in contact with the likely result of a stellar merger.    Due to the practical limitations of the design of our rapid code we cannot follow the radii of stars during the accretion process.  To
account for systems that come in contact as a result of this effect we
 adopt a critical mass ratio for main-sequence binaries, $q_{\rm crit, MS}$, \changes{and assume that the stars come in contact when  $M_{\rm acc}/M_{\rm don} < q_{\rm crit, MS}$, where $M_{\rm don}$ is the mass of the Roche lobe filling star  and $M_{\rm acc}$ is the mass of the companion. }This 
critical mass ratio depends only weakly on the separation \citep{de-Mink+2007}.  We
therefore consider two free parameters $q_{\rm crit, MS}$ and $q_{\rm crit, HG}$ which set the critical mass ratio for systems that interact on the main sequence and systems in which the donor star fills its Roche lobe while it is crossing the Hertzsprung gap.  Our standard value $q_{\rm crit, MS}=0.65$ is based on detailed binary evolutionary models \citep{de-Mink+2007} and for $q_{\rm crit, HG}$ we adopt a value of 0.4. For the advanced evolutionary stages we follow \citet{Hurley+2002}.

\subsubsection {Rejuvenation and the convective core}
If the accreting star is a main-sequence star we assume that it adapts
its interior structure to its new mass. This implies that the convective core grows, such that fresh hydrogen is mixed to the central regions effectively rejuvenating the accreting star.  We treat this by adapting the effective fractional main-sequence life time as in \citet{Tout+1997}.The original implementation implicitly assumed that the
size of the convective core is a fixed fraction of 10\%  the stellar mass. This strongly underestimates the core mass for massive main-sequence stars found in detailed models and ignores the fact that mass fraction of the convective core increases with stellar mass.  To improve this adopt fit formula provided by \changes{\citet{Glebbeek+2008a}} for the effective core mass  the end of the main sequence as a function of the initial stellar mass. We do not account for the shrinkage of the convective core over the course of the main-sequence evolution. 

\subsubsection {Transfer of angular momentum accretion}
When mass is transferred through the interior Lagrange point to the
secondary star is may either form an accretion disk around the
secondary star or it may directly impact on the surface of the
secondary star.  Direct evidence for this is found in the H$\alpha$
profiles in Algol type binaries \citep{Richards+1999}

To distinguish between direct impact and the formation of an accretion
disk we estimate the minimum $R_{\min}$ distance between the stream
and the accreting star using an analytic fit by \citep{Ulrich+1976}
against calculations of \citep{Lubow+1975} to determine whether the
stream will hit the star (Pols, priv. communication).   \citet{Lubow+1975} estimate the specific
angular momentum of the impact stream to be equal to $\sqrt{G M 1.7
  R_{\min}}$.  If the impact parameter is larger then the radius of
the accretor, the stream is assumed to collide with itself after which
viscous process lead to the formation of a Keplerian accretion disk.
The star is assumed to accrete from the inner radius of the disk,
where gas particles accrete with the specific angular momentum of a
Keplerian orbit with a radius equal to the stellar radius $\sqrt{G M
  R_{\rm eq}}$.

\citet{Packet1981} pointed out that, under these assumptions, the
accreting star is very efficiently driven towards the Keplerian limit 
after accreting only a few percent of its own mass in the case of accretion via a disk.  In our standard
models set we will follow the argument of \citet{Paczynski1991} and \citet{Popham+1991} who argue that a star rotating at the Keplerian limit star can keep
accreting mass from an accretion disk without accreting angular
momentum as a result of viscous coupling.
 
While the star rotates near the Keplerian limit the net accreted angular momentum is slowly reduced below the specific angular momentum of a Keplerian orbit as
discussed by \citet{Colpi+1991}. There is some debate whether an equilibrium situation may be reached
at a sub Keplerian rotational velocity, for example in the presence of a
magnetic field \citep{Ghosh+1979,Pringle1989}.  If this equilibrium is reached at a very small fraction of the Keplerian rate it would affect the results in this study, similar as discussed in Sec.\ref{Bfields}.
  
\subsubsection{ Wind accretion}

We consider wind accretion \citep{Bondi+1944} with adopting an
efficiency factor $\alpha_{\rm BH}=1.5$ as detailed in \citet{Hurley+2002}.  However, in contrast to Hurley et al. we assume that the specific angular momentum from material gained by wind accretion is small and can be ignored \citep[cf.][]{Ruffert1999}.

\subsubsection {Treatment of stellar mergers}
A large fraction of massive stars are found in binaries that are
expected to evolve into contact systems and eventually merge
\citep{Benson1970a, Wellstein+2001}.  The mergers that are of main
interest for this study are those that result from the merger of
two main-sequence stars.  We use the extensive grid of detailed binary models
evolutionary models \citet{de-Mink+2007} to calibrate the critical mass ratio for close
binaries leading to contact systems.

For the treatment of mergers involving an evolved star we follow \citet{Hurley+2002}. The original prescription for mergers involving two main-sequence stars assumes that the merger product is completely mixed and that no mass is lost during the process. These assumptions in combination with the original prescription that severely underestimated the mass of the convective core for massive main-sequence stars implied that the mergers experience severe rejuvenation placing them near the zero-age main sequence. 

We adapt the treatment of mergers between two main-sequence stars with
respect to the original implementation to better reflect current
insights and detailed models. We assume that a certain fraction $\mu_{\rm loss}$ of the combined mass is lost from the system when two stars merge.  In our standard simulation we adopt
a fraction of 10\% in our standard simulations in agreement with
\citet{Lombardi+1995,Lombardi+1996}. \changes{Note, however, that these simulations are for direct collisions, not for merging binaries.  We therefore consider $\mu_{\rm loss}$ one of the uncertain parameters}.  We assume that the mass lost from the system originates from  the stellar envelopes and is not enriched in helium.  We assume that, after short
phase lasting a thermal timescale the merger product will settle and
can be described as a stars rotating near the Keplerian rate.
numerical reasons.

Whether merger products experience significant mixing is uncertain \citep[e.g.][]{Gaburov+2008}
 To mimic the result of SPH simulations which show remarkably little
mixing occurring when two stars merge we envision in our standard
model that the core of the most evolved star settles in the
center of the merger surrounded by the core of its companion.  The most evolved star is not necessarily the primary star.  We then assume that the merger product develops a convective core of a mass that can be approximated by the prescriptions by  \citet{Glebbeek+2008a}.  Depending on how much mass is lost from the system, the new convective core mass may be larger or smaller than the  combined mass of both original cores.  

We consider the possibility that an additional amount of mass just above the convective core can be mixed into the central regions.  To this extent we introduce a second parameter  $\mu_{\rm mix}$ which expresses the mass of this region as a fraction of the envelope mass.   We adopt $\mu_{\rm mix} = 0.1 $ as a standard option \citep{Gaburov+2008}.  Setting $\mu_{\rm mix}= 1$ allows for complete mixing of the merger product after mass loss.   We work out the new relative age by considering which regions are mixed and following the approach by \citep{Hurley+2002} to use a simple  linear map between the core composition and the relative age.     

In this new approach, which uses a more realistic approximation for the core size,  the mergers are the relative age of the merger product is only slightly smaller than the relative age of the original primary star.   In other words we now provide a more conservative estimate of the remaining lifetime of the merger products.

\subsubsection{Common envelope evolution}
To treat common envelope situations we follow \citet{Hurley+2002} we
use choose a common envelope ejection efficiency parameter
$\alpha_{\rm CE} = 0.2$ in agreement with studies of binary systems in
planetary nebulae \citep{Zorotovic+2010, Davis+2010, De-Marco+2011}.
For the parameter describing the binding
energy of the envelope $\lambda_{\rm CE}$ we use fits to
\citep{Dewi+2000} and we adopt  $\lambda_{\rm ionization} = 0.5$.
We note that in the original version of the code massive stars  are not treated 
as convective giants when they become red super giants after the ignition of helium. We corrected this using the base of the giant branch as a transition point.
During common envelope evolution we assume that the companion does not
accrete mass nor angular momentum.  Our results are not sensitive to these assumptions.

%\bibpunct{[}{]}{,}{a}{}{;}
\bibliography{my_bib}

\end{document}